\documentclass{ws-aada}

\usepackage{amsmath}
\usepackage{subfigure}
\usepackage{float}
\usepackage{epsfig}
\usepackage{color}

\begin{document}

%
\catchline{}{}{}{}{}
%
 
\markboth{D.N. Kaslovsky and F.G. Meyer}{Noise Corruption of EMD and Instantaneous Frequency}

\title{NOISE CORRUPTION OF EMPIRICAL MODE DECOMPOSITION \\AND ITS EFFECT ON INSTANTANEOUS FREQUENCY} 

\author{DANIEL N. KASLOVSKY}
\address{Department of Applied Mathematics, University of Colorado at Boulder, 526 UCB\\
Boulder, Colorado 80309-0526, USA\\
\email{kaslovsky@colorado.edu}
}

\author{FRAN\c{C}OIS G. MEYER}
\address{Department of Electrical Engineering, University of Colorado at Boulder, 425 UCB\\
Boulder, Colorado 80309-0425, USA\\
fmeyer@colorado.edu}


\maketitle

\begin{abstract}
Huang's Empirical Mode Decomposition (EMD) is an algorithm for analyzing nonstationary data that provides a localized time-frequency representation by decomposing the data into adaptively defined modes.  EMD can be used to estimate a signal's instantaneous frequency (IF) but suffers from poor performance in the presence of noise.  To produce a meaningful IF, each mode of the decomposition must be nearly monochromatic, a condition that is not guaranteed by the algorithm and fails to be met when the signal is corrupted by noise.  In this work, the extraction of modes containing both signal and noise is identified as the cause of poor IF estimation.  The specific mechanism by which such ``transition'' modes are extracted is detailed and builds on the observation of Flandrin and Goncalves that EMD acts in a filter bank manner when analyzing pure noise.  The mechanism is shown to be dependent on spectral leak between modes and the phase of the underlying signal.  These ideas are developed through the use of simple signals and are tested on a synthetic seismic waveform.
\end{abstract}


\section{Introduction}

Empirical Mode Decomposition (EMD) is an analysis tool for nonstationary data introduced by \citeauthor{HuangOrig} in 1998.  Nonstationary signals have statistical properties that vary as a function of time and should be analyzed differently than stationary data.  Rather than assuming that a signal is a linear combination of predetermined basis functions, the data are instead thought of as a superposition of fast oscillations onto slow oscillations \cite{Flandrin}.  EMD identifies those oscillations that are intrinsically present in the signal and produces a decomposition using these modes as the expansion basis.  We note that throughout this work the term ``basis'' is used in the same sense as used by Huang \shortcite{HuangOrig}: the modes of a signal's decomposition do not span a particular space, but provide an expansion for the specific signal.  In this way, the basis is data driven and adaptively defined each time a decomposition is performed \cite{Flandrin}.  EMD has been used for data analysis in a variety of applications including engineering, biomedical, financial and geophysical sciences \cite{HHT}.\\

In contrast with Fourier analysis, EMD requires no assumptions on its input and is therefore well suited to analyze nonstationary data.
Since nonstationarity implies that a signal is not well represented by pure tones, a significant number of harmonics is required to represent a nonstationary signal in the Fourier basis.  Energy must be spread across many modes to accommodate deviations from a pure tone.  In producing an adaptive decomposition consisting of modes that allow for such deviations, EMD efficiently represents the signal by relaxing the need to explore all frequencies.  A signal is expanded using only a small number of adaptively defined modes.\\

As EMD is an algorithm and not yet a theoretical tool, its limits must be tested experimentally.  Several authors have reported on its performance in the presence of noise \cite{HHT, Flandrin, ND}.  In this work, we propose a new understanding of the mechanism that prevents the algorithm from properly estimating the instantaneous frequency of a noisy signal.  The paper is organized as follows.  Section 2 gives a brief description of the EMD algorithm and demonstrates its use in estimating the instantaneous frequency of a clean signal.  The same estimation, performed in the presence of noise, is seen to be problematic in Section 3 and the cause is identified.  Section 4 outlines a new explanation for this poor performance.  Finally synthetic seismic data are used in Section 5 to extend our study from simple signals to a model for real world data.\\

\section{Empirical Mode Decomposition}

\subsection{Algorithm}
The goal of Empirical Mode Decomposition is to represent a signal as an expansion of adaptively defined basis functions with well defined frequency localization. Each basis function, called an Intrinsic Mode Function (IMF), should be physically meaningful, representing ideally one frequency (nearly monochromatic).  To accomplish this, an IMF is defined as a function for which (1) at any point, the mean of the envelopes defined by local maxima and minima is zero, and (2) the number of extrema and the number of zero crossings differ by at most one \cite{HuangOrig}.  Such a definition attempts to ensure that a meaningful instantaneous frequency can be obtained from each IMF, a process that is defined and detailed in the next subsection, but does not guarantee that each IMF is narrow band \cite{HuangOrig}.  To decompose a signal $x(t)$, the EMD algorithm works as follows \cite{Flandrin}:
\begin{enumerate}
	\item Interpolate (usually with cubic splines) the local maxima of $x(t)$ to form an envelope.  Repeat for the minima.
	\item Compute the mean, $m(t)$, of the two envelopes.
	\item Compute the detail, $d(t)$, by subtracting the mean from the signal, $d(t) = x(t) - m(t)$.  Extract the detail as an IMF.
	\item Repeat the iteration on the residual $m(t)$.  Continue until the residual is such that no IMF can be extracted and represents the trend.
\end{enumerate}
While the trend does not meet the definition of an IMF, we will refer to it as the final IMF for convenience.  Before the detail, $d(t)$, can be considered an IMF, a ``sifting'' process takes place during which the detail is treated as a new signal and is iterated until a predefined stopping criterion is reached.  The purpose of this step is to enforce the definition of an IMF \cite{Flandrin}.  Ideally, all modes are now nearly monochromatic and can be used to give a meaningful estimate of the signal's instantaneous frequency.\\

The algorithm can be described in the time-frequency domain as a collection of data-dependent projections.  \citeauthor{Olhede} \shortcite{Olhede} formalize this idea by defining projection operators $P_{R_{j}}$, not necessarily orthogonal, that project a signal $x(t)$ into regions $R_j$ of the time-frequency plane.  The signal may then be written as
\begin{equation*}
	x(t) = \sum_{j=1}^{K} [P_{R_{j}}x](t),
\end{equation*}
where $K$ is the number of IMFs produced, with the $K$th IMF being the trend.  Since each projection gives rise to an IMF, an expansion of the signal is then given by
\begin{equation*}
	x(t) = \sum_{j=1}^{K} X_j(t),
\end{equation*}
where $X_j$ is the $j$th IMF.\\

\subsection{Estimation of instantaneous frequency}
A signal is often characterized in terms of its frequency content.  When a signal's statistical properties are shift-invariant in time, it is said to be stationary.  As this definition implies, frequency remains constant throughout the signal's duration, and is easily defined as the number of periods per unit time.  However, if the signal's frequency varies with time, it is said to be nonstationary, and this global definition of frequency loses meaning.  It is therefore necessary to characterize the frequency content of the signal in a local manner.  For example, a chirp with a quadratic phase has frequency that changes linearly from one instant to the next.  It is not possible to pinpoint one frequency for the entire chirp.  Instead the chirp's frequency is described as a (linear) function of time.  It is therefore more useful to characterize such a signal in terms of its instantaneous frequency.\\

Boashash \shortcite{Boashash1} describes instantaneous frequency (IF) as ``a time-varying parameter which defines the location of the signal's spectral peak as it varies with time.''  He points to seismic, radar, sonar, communication, and biomedical applications as fields where IF is utilized.  Two conditions are needed to produce a physically meaningful and well defined instantaneous frequency.  The signal must be analytic and it must be narrow band.  An analytic signal is produced via the Hilbert transform:
\begin{equation*}
	[\mathcal{H}x](t) = \frac{1}{\pi}PV\int_{-\infty}^{\infty}\frac{x(t')}{t-t'}dt',
\end{equation*}
where PV denotes the Cauchy principle value.  Given a real valued signal, $x(t)$, its analytic representation is then defined as $z(t) = x(t) + i[\mathcal{H}x](t).$  The analytic signal $z(t)$ may be written in the form
\begin{equation*}
 	z(t) = a(t)e^{i\phi(t)},
\end{equation*}
and the instantaneous frequency, $v(t)$, can then be defined \cite{Boashash1} in terms of the derivative of the phase $\phi(t)$:
\begin{equation*}
	v(t) = \frac{1}{2\pi}\frac{d\phi}{dt}.
\end{equation*}
The derivative must be well defined since physically there can only be one instantaneous frequency value $v(t)$ at a given time $t$.  This is ensured by the narrow band condition: the signal must contain nearly one frequency.  Further, as detailed by \citeauthor{Boashash1} \shortcite{Boashash1}, the Hilbert transform produces a more physically meaningful result the closer its input signal is to being narrow band.  However, we wish to work with signals that are much more interesting than those that are monochromatic.  This can be achieved by decomposing such a signal into several nearly monochromatic components, each of which provides a well defined, meaningful instantaneous frequency.  An overall IF estimate of a signal $x$, given its decomposition into $K$ IMFs, is then calculated as a weighted sum of the individual IFs:  
\begin{equation*}
	IF\bigl(x(t)\bigr) = \frac{\sum_{j=1}^{K} A_j^2(t)v_j(t)}{\sum_{j=1}^{K} A_j^2(t)},
\end{equation*}
where $A_j(t)$ and $v_j(t)$ are, respectively, the magnitude and instantaneous frequency of the analytic representation of IMF $X_j$ \cite{Olhede}.\\

\begin{figure}[H]
\centerline{
\subfigure[IF estimate]{\psfig{file=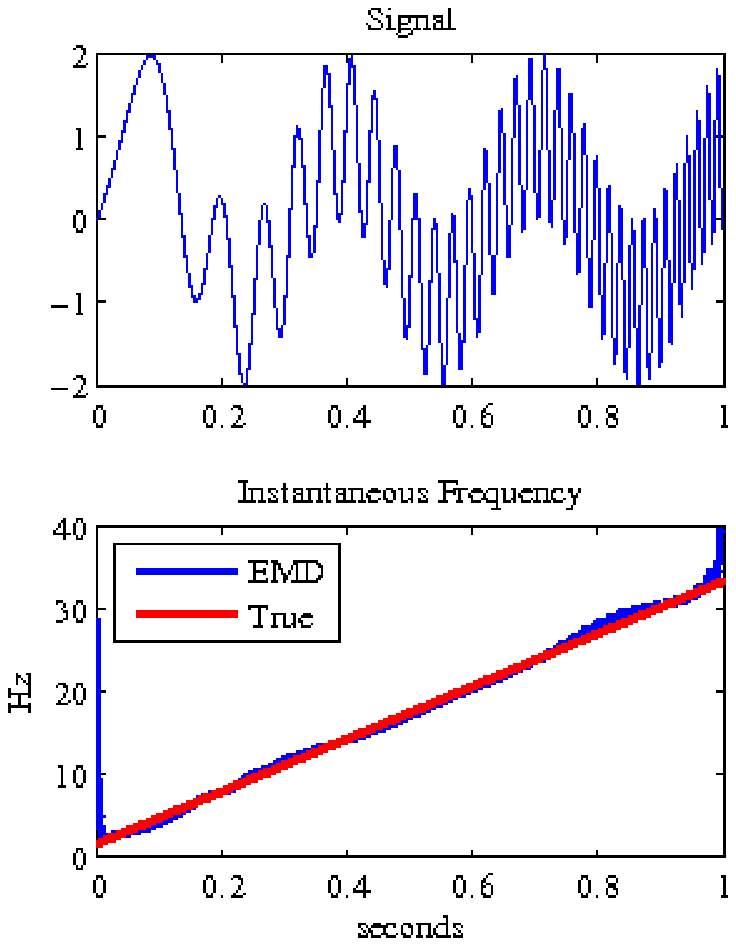}}
\subfigure[IMFs]{\psfig{file=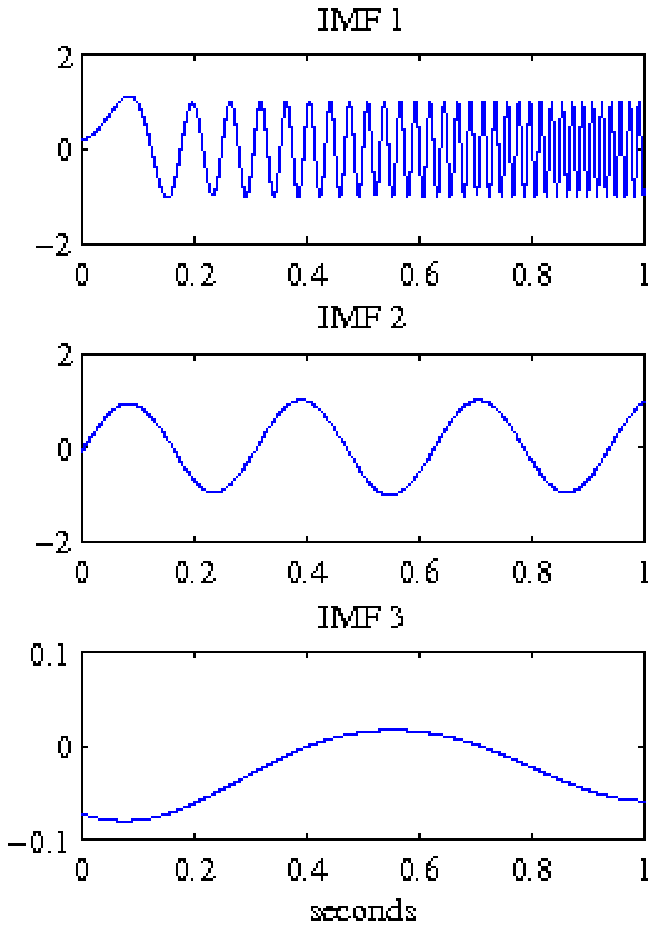}}
}
\vspace*{8pt}
\caption{The instantaneous frequency estimate and IMFs of a clean signal.}
\label{cleanIF}
\end{figure}

To demonstrate the calculation of IF, consider $x(t) = \sin(200t^2) + \sin(20t)$, the superposition of a linear chirp onto a stationary sine wave, on the interval $t \in [0,1]$ seconds.  Figure \ref{cleanIF}a shows the true analytic\footnote{The analytic IF of the superposition of two signals, $x(t) = A_1(t)e^{i\phi_1(t)}+A_2(t)e^{i\phi_2(t)}$, is defined as the average of the individual IFs of each signal only when $|A_1(t)| = |A_2(t)|$ \cite{LoughlinTacer}.  We note that this condition holds for this example, and we compute the analytic IF accordingly.  An example for which the condition does not hold will be encountered in section 5.} IF (in red) and the overall IF estimate (in blue) obtained from the IMFs (shown in figure \ref{cleanIF}b) of $x(t)$.  We are able to calculate a physically meaningful instantaneous frequency when using the decomposition of a signal in the absence of noise.\\

\citeauthor{onIF} \shortcite{onIF} give a detailed discussion on the shortcomings of this method of IF calculation.  In particular, they note that the analytic signal obtained from the Hilbert transform is only physically meaningful if the conditions of the Bedrosian theorem are met.  They introduce a normalization scheme that empirically separates the AM and FM components of each IMF, where the AM carries the envelope and the FM is the constant amplitude variation in frequency.  The ``normalized'' FM component of an IMF is guaranteed to satisfy the Bedrosian theorem and is therefore suitable for the Hilbert transform.
Alternatively, once an IMF has been normalized, \citeauthor{onIF} \shortcite{onIF} propose eschewing any Hilbert transform in favor of applying a $90$ degree phase shift by means of a direct quadrature.  Both methods are demonstrated to be more accurate on clean signals than the standard method presented above.  Since the focus of our work is the performance of EMD in the presence of noise, the performance of this normalization scheme on noisy data will be addressed in the next section.\\


\section{Performance in the Presence of Noise}

A clean signal can produce a decomposition that lends itself to a meaningful instantaneous frequency estimate.  However, as is the case in many applications, data are often contaminated by noise.  Decomposing a noisy signal produces both narrow and wide band IMFs.  While most of the wide band IMFs contain noise and may be discarded, a small number capture the transition from noise to signal and must be kept.  This leads to a corrupted estimate of the instantaneous frequency.\\

\subsection{Evidence of a problem}
\begin{figure}[H]
\centerline{\psfig{file=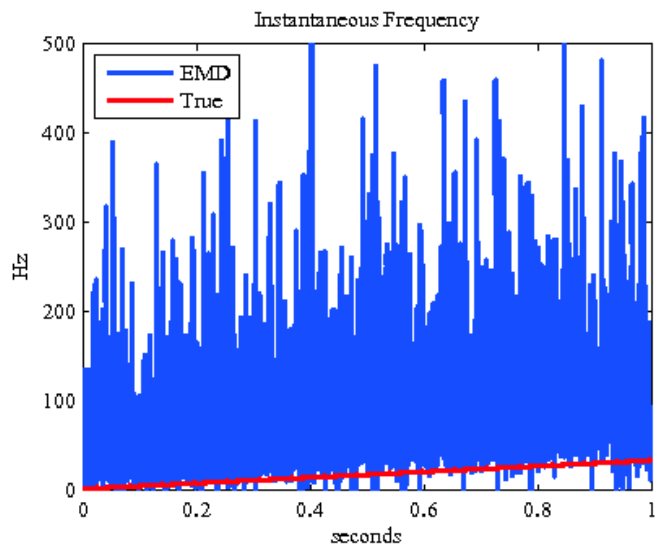,clip,trim=2in 3.7in 2in 3.85in,scale=.85}}
\vspace*{8pt}
\caption{The corrupted instantaneous frequency estimate of a noisy signal.}
\label{noisyIF}
\end{figure}

In the previous section the calculation of instantaneous frequency was described.  This process is now applied to the same signal contaminated with additive white Gaussian noise such that its SNR is 27dB.  Throughout this work we use SNR $= 10\log_{10}\bigl(\frac{\|x\|_2}{\sigma}\bigr)$dB, where $\sigma$ is the standard deviation of the noise.  The result is shown in figure \ref{noisyIF} and it is clear that a meaningful instantaneous frequency estimate was not produced.
\noindent To understand this poor result, recall that a signal's IF is computed as a weighted sum of the IF from each of its IMFs.  The analytic representation of each IMF is required and thus each IMF must be narrow band to ensure a meaningful Hilbert transform.  Moreover, IF is well defined only in the case of a nearly monochromatic signal.  Therefore, for the purpose of computing a meaningful IF, the key feature of the decomposition is that each IMF contains nearly one frequency.\\

\begin{figure}[H]
\centerline{\psfig{file=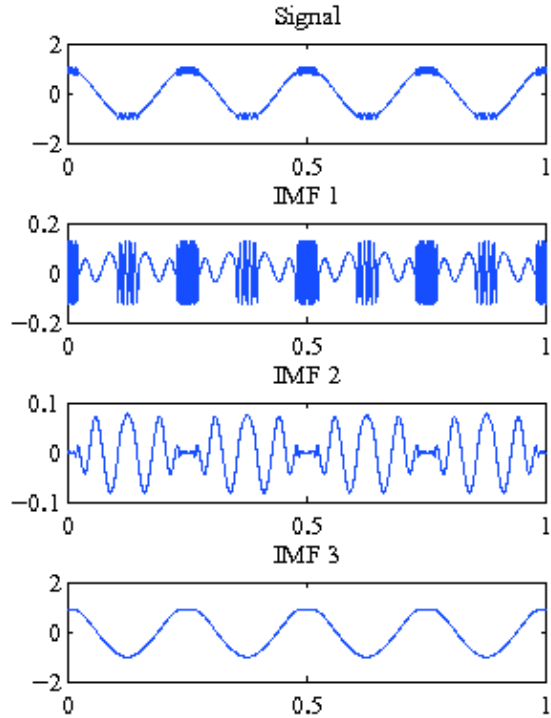,clip,trim=2in 3.6in 2in 3.5in}}
\vspace*{8pt}
\caption{IMFs of a deterministic signal.  IMFs 1 and 2 contain both high and low frequencies, illustrating that monochromaticity is not guaranteed.}
\label{breakEMD}
\end{figure}

It is important to recall that the definition of an IMF does not guarantee monochromaticity.  This is illustrated with a deterministic example.  The decomposition of a signal composed of a slow sinusoid with high frequency sinusoids superimposed at each crest and trough is shown in figure \ref{breakEMD}.  Despite the fact that this signal was constructed in a completely deterministic manner, its first two IMFs contain both high and low frequencies.  Such IMFs are not suitable for the Hilbert transform and will not yield a well defined IF.  \citeauthor{WuHuang} \shortcite{WuHuang} use a very similar example, developed independently from our example, to note that a decomposition may give rise to IMFs containing oscillations of drastically different scales.  They refer to the creation of such IMFs as ``mode mixing,'' and introduce the Ensemble EMD (EEMD) to alleviate this issue.  We will discuss the performance of EEMD on noisy data in section 4.\\

\subsection{Identifying the culprit}
\begin{figure}[H]
\centerline{\psfig{file=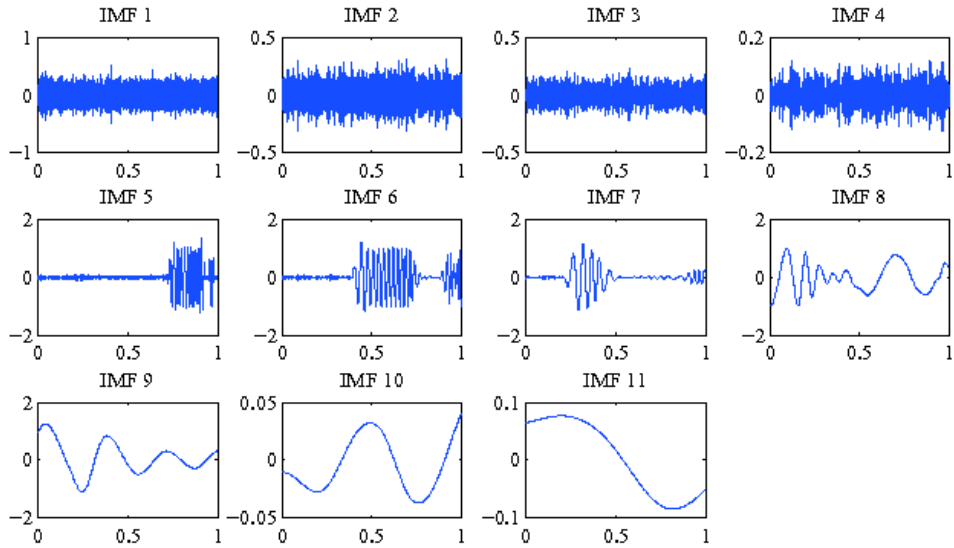,clip,trim=1in 3.5in 1in 3.5in,scale=.8}}
\vspace*{8pt}
\caption{IMFs of a noisy signal.  IMFs 1-4 capture most of the noise, while IMFs 5-7 represent the transition from noise to signal, and IMFs 8-11 are nearly monochromatic.}
\label{noisy allIMFs}
\end{figure}

The poor quality IF estimate from a noisy signal can be explained by the creation of wide band IMFs.
More precisely, the EMD decomposition of a noisy signal will generate some ``noisy'' IMFs.  As explained below, such noisy IMFs are neither monochromatic signals nor pure noise; rather their Fourier transform is localized over a well defined frequency range.  Consequently, such IMFs cannot contribute a well defined IF because noise is wide band by definition.  Figure \ref{noisy allIMFs} shows the decomposition of the noisy example signal.  We identify three categories of IMFs:\\
\begin{enumerate}
	\item \textbf{Noisy}: IMFs 1-4 are wide band as they clearly contain noise.
	\item \textbf{Transition}: IMFs 5-7 contain both signal and noise.  These IMFs capture the ``transition'' from the noise captured in IMFs 1-4 and the monochromatic components extracted as IMFs 8-11.
	\item \textbf{Monochromatic}: IMFs 8-11 are nearly monochromatic and yield meaningful IF contributions.
\end{enumerate}

\begin{figure}[H]
\centerline{
\subfigure[IMF 2 - Noisy]{\psfig{file=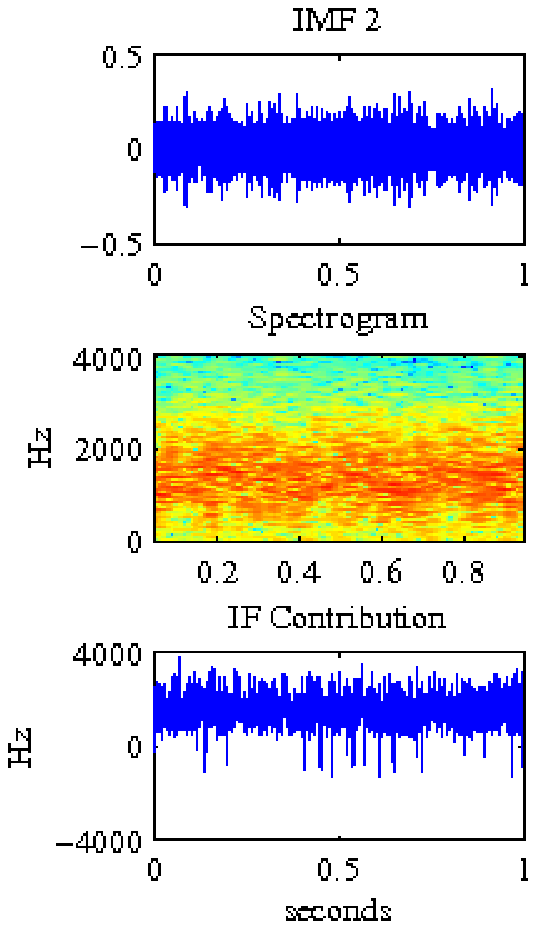}}
\subfigure[IMF 5 - Transition]{\psfig{file=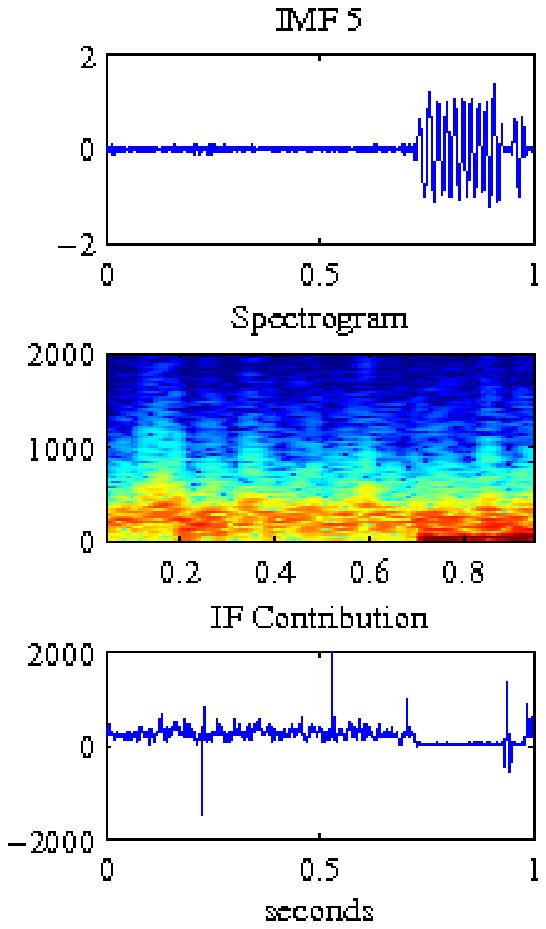}}
\subfigure[IMF 9 - Monochromatic]{\psfig{file=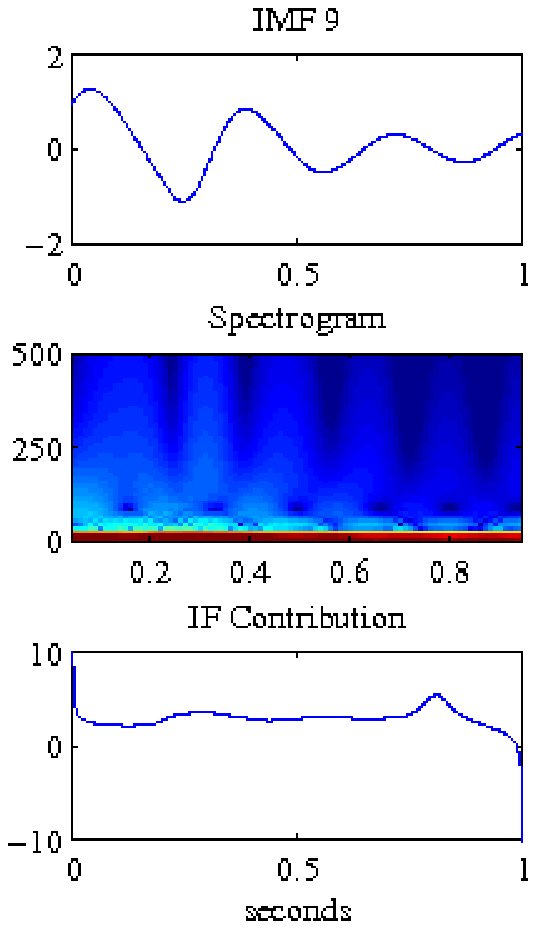}}
}
\vspace*{8pt}
\caption{Characteristic IMFs representing (a) noise, (b) transition from noise to signal, and (c) monochromatic components extracted from a noisy signal.}
\label{exIMFs}
\end{figure}

To demonstrate the effect of each type of IMF on the overall IF estimate, figure \ref{exIMFs} highlights an example from each category.  IMF 2 (left) is a noisy IMF; IMF 5 (center) contains both signal and noise and is a transition IMF; IMF 9 (right) is nearly monochromatic.  Spectrograms\footnote{Spectrograms are displayed as a log-scale color representation of the power spectral density calculated using a Kaiser window of duration 0.1 seconds with 90\% overlap.  Red and blue correspond to higher and lower density, respectively, and the scale is uniform within a figure but not necessarily throughout the paper.} are used to illustrate the frequency content that characterizes each IMF.  The spectrogram of the noisy mode, IMF 2, shows that it is wide band and therefore yields an IF that is not physically meaningful.  In contrast, the nearly monochromatic IMF 9 is seen to be narrow band and contributes a well defined IF.  Finally, despite its signal content, transition IMF 5 is wide band and cannot contribute a clean IF.  The inclusion of IF contributions from wide band IMFs pollutes the overall IF and is responsible for the poor result seen in figure \ref{noisyIF}.\\

\begin{figure}[H]
\centerline{\psfig{file=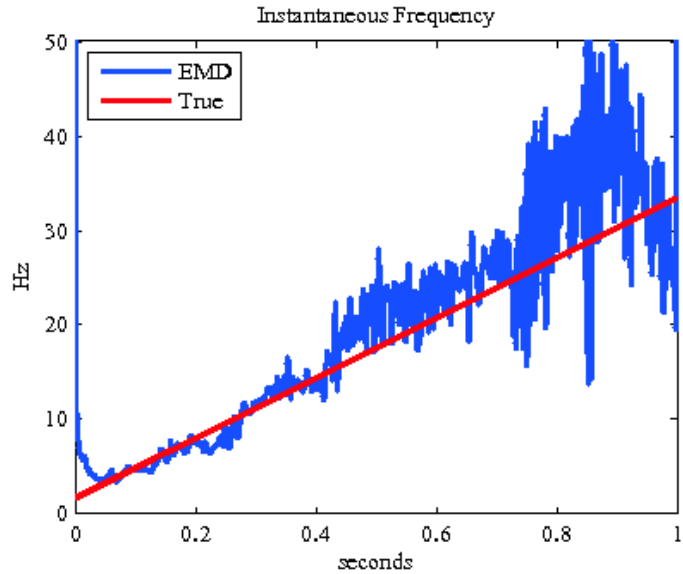,clip,trim=2in 3.6in 2in 3.85in,scale=.9}}
\vspace*{8pt}
\caption{Instantaneous frequency estimate using IMFs 5-11.  The necessary inclusion of transition IMFs prevents a clean estimation.}%
\label{cleanedupIF}%
\end{figure}

Since the inclusion of certain IMFs results in a poor IF estimate, it is reasonable that some nonlinear thresholding process would yield better results.  Specifically, discarding the IMFs identified as noise will provide a more meaningful IF estimate.  In figure \ref{cleanedupIF} the IF of the noisy version of the signal shown in figure \ref{cleanIF}a(top) is now computed using only IMFs 5-11.  It is important to note that IMF 5 is not discarded because as a transition IMF, it contains both signal and noise.  We would like to ignore such an IMF since it will provide poor IF information derived partially from noise, but cannot discard its signal content.  Therefore, it must be included and contaminates our overall estimate.  The same is true of IMFs 6 and 7.  Other thresholding methods may be utilized, including using only those IMFs with energy between specified thresholds \cite{HHT}.  However, to our knowledge, there is not a clear cut method of thresholding that will produce a faithful IF estimation.  While the thresholded estimate in figure \ref{cleanedupIF} is an improvement over the previous estimate shown in figure \ref{noisyIF}, the transition IMFs' contribution has left the IF mostly incoherent.  The necessary inclusion of transition IMFs is therefore identified as the main problem in estimating the IF in the presence of noise.\\

\begin{figure}[H]
\centerline{
\subfigure[IMF 2 - Noisy]{\psfig{file=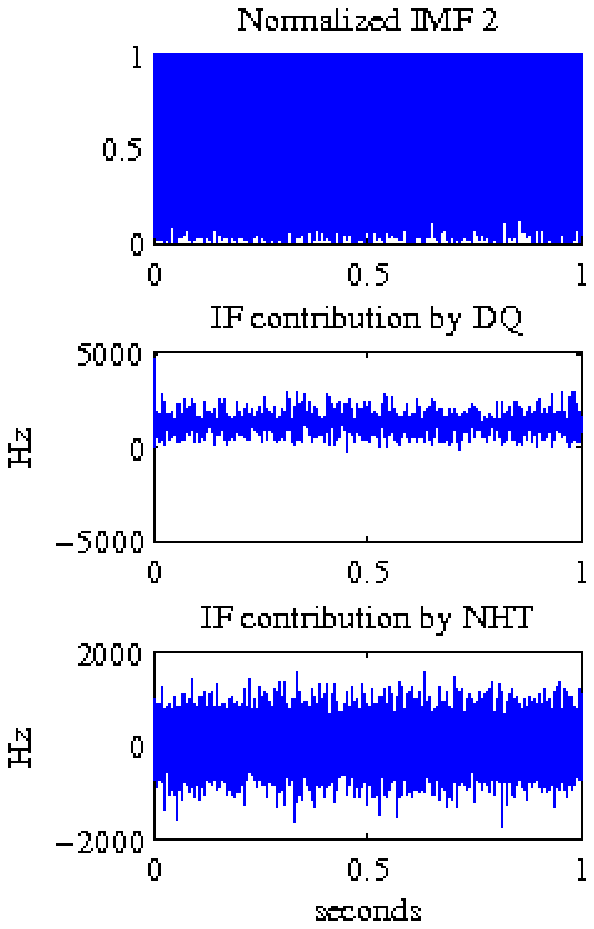}}
\subfigure[IMF 5 - Transition]{\psfig{file=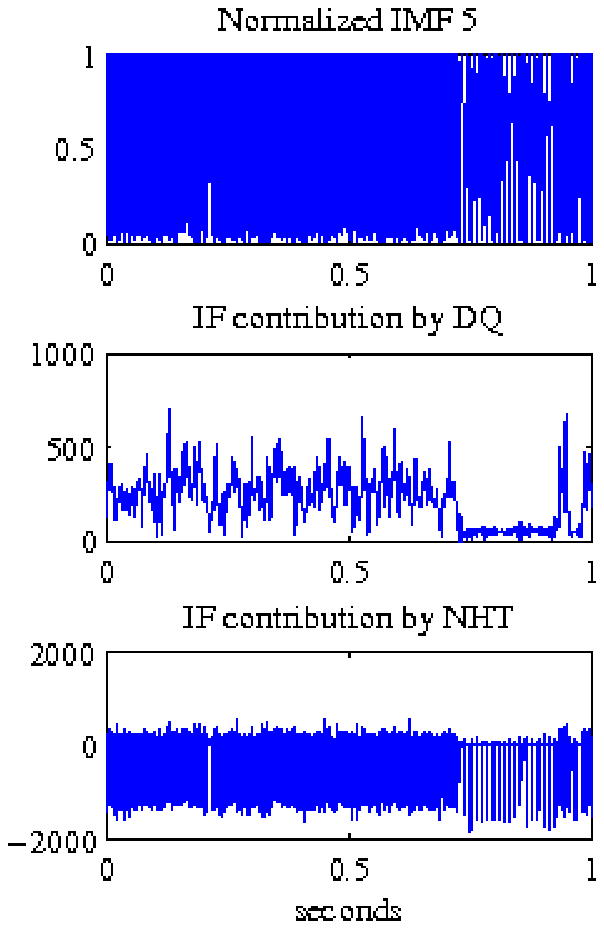}}
\subfigure[IMF 9 - Monochromatic]{\psfig{file=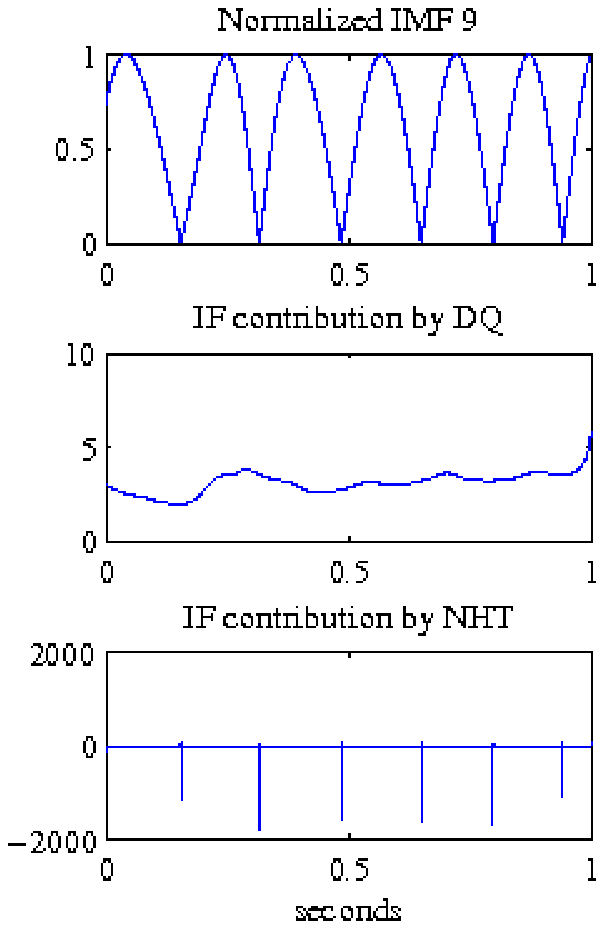}}
}
\vspace*{8pt}
\caption{Normalized IMFs of a noisy signal (top), IF contribution from direct quadrature (middle), and IF contribution from normalized Hilbert transform (bottom).}
\label{normalized-exIMFs}
\end{figure}

It is also reasonable that computing the IF from normalized IMFs \cite{onIF} (see section 2) might yield cleaner results.  However, \citeauthor{onIF} \shortcite{onIF} note that the normalized scheme encounters problems when an IMF contains noise and recommend computing the analytic signal with the standard Hilbert transform approach.  Figure \ref{normalized-exIMFs} shows the normalized version of the example IMFs from figure \ref{exIMFs}.  We observe that we still have (from left to right) a noisy IMF, a transition IMF, and a monochromatic IMF.  The IF contribution from each normalized IMF is shown, calculated by direct quadrature (middle) and normalized Hilbert transform (bottom).  Just as in the standard unnormalized case, transition IMFs with corrupted IF contributions still exist and their necessary inclusion will prevent a clean IF estimate (not shown).\\

\section{Analysis of Noisy Decompositions}

With an understanding of how transition IMFs pollute the estimation of IF, we address the more fundamental question of why transition IMFs are produced when EMD operates on a noisy signal.  To begin, we note the work of Flandrin and Goncalves \shortcite{Flandrin} showing that EMD acts as a filter bank when decomposing pure noise, and add our observation that the boundaries of the frequency bands vary with time.  We propose two mechanisms that lead to the creation of transition IMFs:
\begin{enumerate}
	\item \textbf{Spectral leak} between frequency bands: frequency content of the underlying signal falls within a band treated as noise.
	\item \textbf{Phase alignment}: the alignment of the signal with the lowest level of noise present in the band is controlled by the signal's phase.
\end{enumerate}
Spectral leak is mostly a nonstationary condition while the contribution of phase alignment is best seen in the stationary setting.

\subsection{EMD decomposition of pure white noise}
\begin{figure}[H]
\centerline{\psfig{file=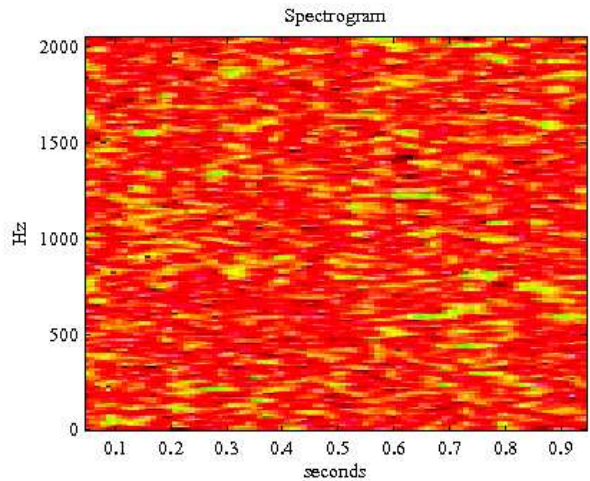,clip,trim=2in 3.6in 2in 3.85in,scale=.75}}
\vspace*{8pt}
\caption{Spectrogram of white Gaussian noise used throughout this section.}%
\label{noise}%
\end{figure}

Before returning to the decomposition of a noisy signal, EMD's performance on pure noise is analyzed.  Figure \ref{noise} shows the spectrogram of a realization of white Gaussian noise (zero mean, standard deviation of 0.2).  It is not surprising that the spectrogram shows nearly uniform power spectral density since, in principle, the density of such noise should be constant.  This specific noise realization will be used in all experiments that follow in this section.\\
 
\begin{figure}[h]
\centerline{\psfig{file=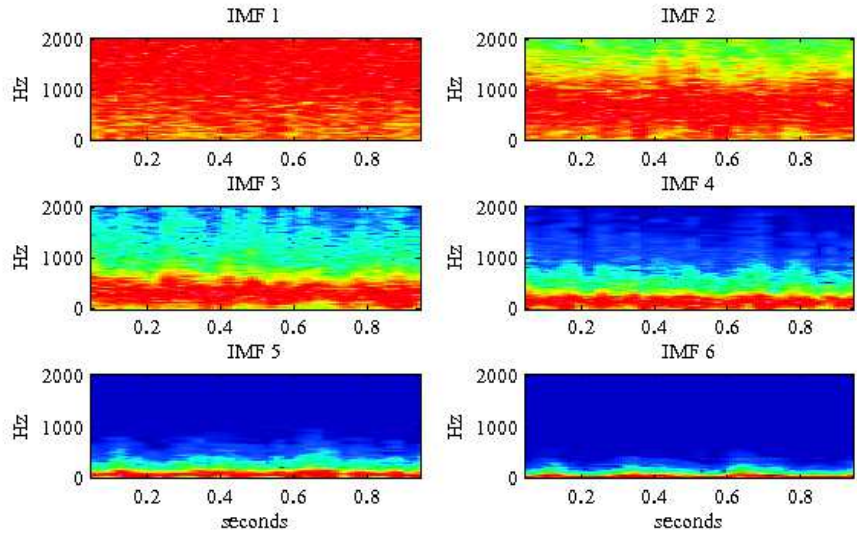,clip,trim=1in 3.7in 1in 3.7in,scale=.8}}
\vspace*{8pt}
\caption{Spectrogram of first six IMFs of white Gaussian noise, highlighting EMD's filter bank behavior.}
\label{noiseIMFsSpec}
\end{figure}

\begin{figure}[H]
\centerline{\psfig{file=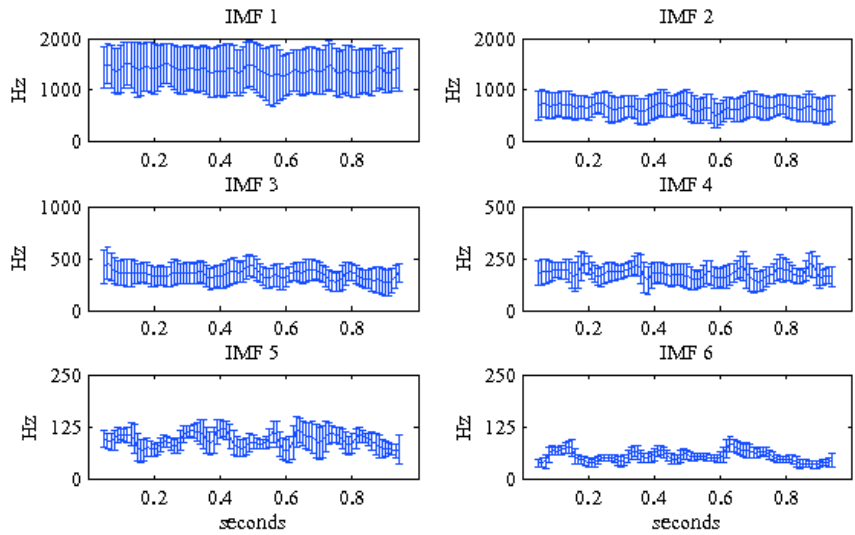,clip,trim=1in 3.7in 1in 3.7in,scale=.8}}
\vspace*{8pt}
\caption{Mean (with error bars representing one standard deviation) power spectral density of IMFs extracted from white Gaussian noise.  Note the different scales on the frequency axis, clearly indicating an almost dyadic decomposition of the noise spectrum.}
\label{noiseIMFsStat}
\end{figure}

Flandrin and Goncalves \shortcite{Flandrin} reported that EMD acts as a filter bank when decomposing pure Gaussian noise.  By selecting entire frequency bands as IMFs rather than a single frequency, the IMFs are by definition multicomponent.  We observe a similar result and note that the boundaries of each band are not straight line cuts through the frequency axis, but instead vary as a function of time.
This is clearly seen in the IMFs of the noise as their spectrograms (figure \ref{noiseIMFsSpec}) show that the borders of the frequency-bands do not resemble straight lines.  The spectrograms also reveal that the IMFs provide a nearly dyadic decomposition of the spectrum shown in figure \ref{noise}.  Since the noise is composed of realizations of random variables, we define its mean power spectral density $M_{psd}(t)$ and associated standard deviation $SD_{psd}(t)$ at a given time $t$ as follows:
\begin{align*}
	M_{psd}(t) &= \sum_{k=0}^{F_s/2} k \cdot P(k,t)\\
	M^2_{psd}(t) &= \sum_{k=0}^{Fs/2} k^2 \cdot P(k,t)\\
	SD_{psd}(t) &= \sqrt{M^2_{psd}(t)-\bigl(M_{psd}(t)\bigl)^2}
\end{align*}
\noindent where $F_s$ is the sampling rate and $P(k,t)$ is the normalized power spectral density at frequency $k$ and time $t$.  The plots of the mean power spectral density with error bars representing one standard deviation show that the statistics of the IMFs vary with time (figure \ref{noiseIMFsStat}).  Some frequency mixing between modes is also observed.\\
\newpage

\subsection{EMD decomposition of a signal corrupted by noise}
\subsubsection{Spectral leak}
Kijewski-Correa and Kareem \shortcite{ND} attributed the poor quality of IF estimation in the presence of noise to the empirical nature of the algorithm, leading to a basis derived from the noise.  They observed the mixing of the input signal over many IMFs, making it difficult to isolate the clean signal from the noise.  We extend this explanation with our observations to explain the extraction of transition IMFs.  The process is best understood by considering the noisy signal in the time-frequency plane.  The algorithm is operating on projections in this plane, starting with the highest frequency band and adaptively moving down the frequency axis.  These projections are not completely orthogonal, and thus there is some frequency mixing in the modes.  As EMD tiles down the time-frequency plane, it first extracts pure noise as it has not yet reached the frequency of the signal.  While in the pure noise region, EMD behaves as a filter bank, as observed by Flandrin and Goncalves, extracting noise in an almost dyadic manner.\\
\begin{figure}[h]
\centerline{\psfig{file=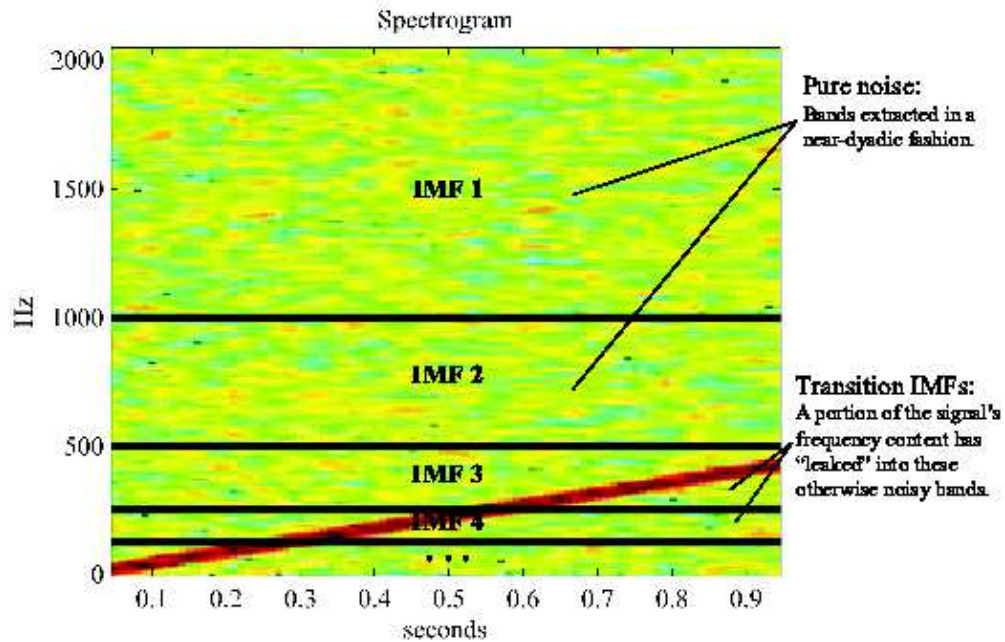}}
\vspace*{8pt}
\caption{A model of EMD's filter bank action shown in the time-frequency plane.  Pieces of chirping signal are captured in noisy bands.  The bands contributing to IMFs 1-4 are illustrated and the boundaries between the bands are idealized.}%
\label{heuristic}%
\end{figure}

A model for this process in the time-frequency plane is provided in figure \ref{heuristic}.  The model shows a spectrogram of the noisy chirp $\sin(2\pi ft^2) + n(t)$, where $f = 225$ Hz, $t \in [0,1]$, and $n(t)$ is the exact same realization of noise shown in figure \ref{noise}.  The boundaries between the bands are idealized, highlighting EMD's filter bank behavior.  Noise is removed until a frequency present in the signal matches or exceeds that of the noise.  The model demonstrates the situation where a portion of a nonstationary signal leaks into an otherwise noisy band (IMF 3 in this example).  In this case, the signal's frequency is high enough to be included in the IMF for only part of its duration.  Still behaving in the noise regime, EMD extracts both signal and noise as it cannot distinguish which should be removed.  Because of the variation in the boundaries of the identified frequency bands (seen in figures \ref{noiseIMFsSpec} and \ref{noiseIMFsStat}, not shown in the model), EMD will encounter such a band even when decomposing a noisy stationary signal.  This is the general process that leads to the creation of a transition IMF, and will be seen explicitly in the following example.\\
\begin{figure}[H]
\centerline{\psfig{file=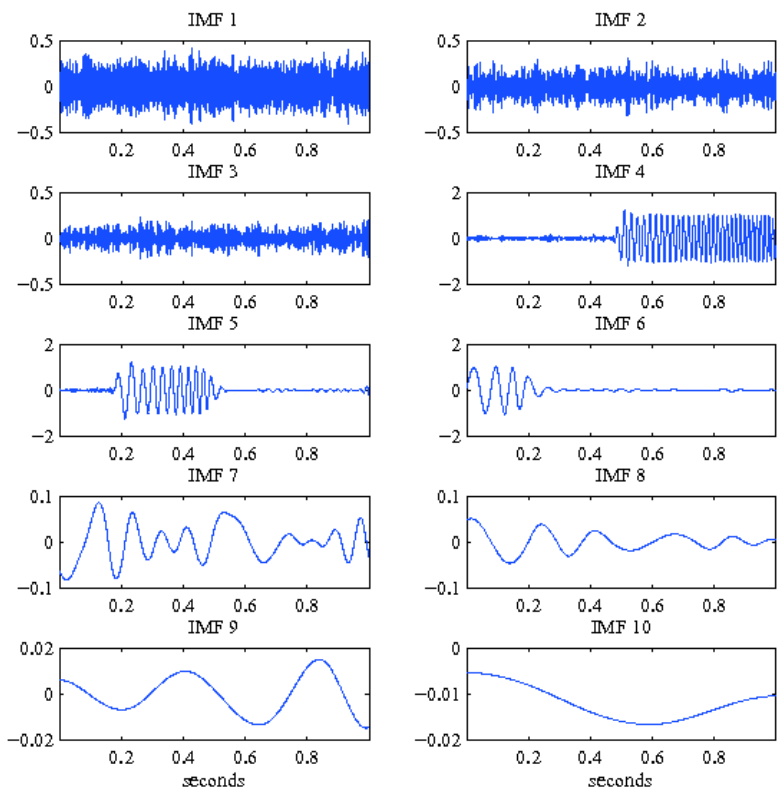,clip,trim=1in 2.75in 1in 2.8in,scale=.75}}
\vspace*{8pt}
\caption{Decomposition of a noisy linear chirp.  Note the signal content present in the transition IMFs 4-6.}
\label{linearChirpIMFs}
\end{figure}

\begin{figure}[H]
\centerline{\psfig{file=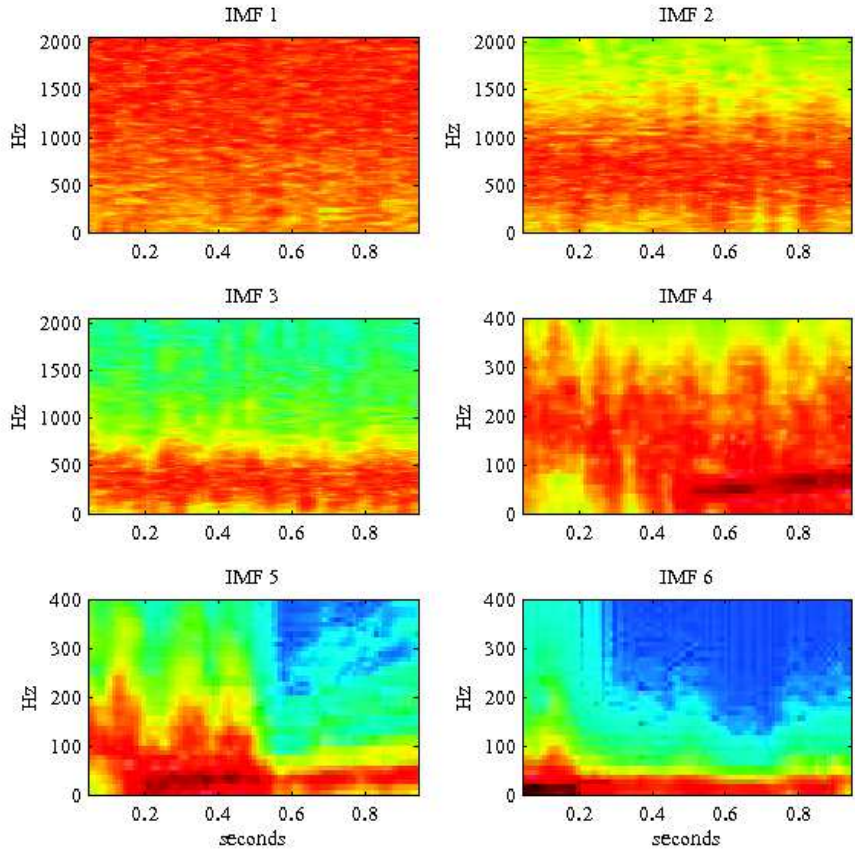,clip,trim=1in 2.75in 1in 2.7in,scale=.8}}
\vspace*{8pt}
\caption{Spectrograms of the decomposition of a noisy linear chirp.  Transition IMFs 4-6 display the spectral leak of signal into noise.  Note the change in scale on the frequency axis.}
\label{linearChirpIMFsSpec}
\end{figure}

To demonstrate the extraction of transition IMFs, we add the exact same realization of noise shown in figure \ref{noise} to the linear chirp $\sin[2\pi (35t^2+10t)]$.  The decomposition of this noisy signal is shown in figure \ref{linearChirpIMFs} and spectrograms of the first six IMFs are shown in figure \ref{linearChirpIMFsSpec}.  IMFs 1-3 show the filter bank action of EMD.  The frequency of the signal is well below that of the noise, and EMD extracts the noise in a nearly dyadic fashion.  We note the boundaries of the frequency bands vary with time, as expected.  Once IMFs 1-3 have been removed, the next frequency band selected contains both noise and signal as can be seen in the spectrogram of IMF 4 (see figure \ref{linearChirpIMFsSpec}).  The noise remaining in the residual forces EMD to continue behaving as a filter bank.  However, the highest frequency content of the chirp now falls within this band.  In removing this band, a portion of the signal is pushed into IMF 4.  In this respect, we observe the signal leaking into the noise.  IMF 4 will be composed of a mixture of noise and signal: noise for the temporal locations corresponding to those where the chirp's frequency is too low to be included; signal for the temporal locations where the chirp's frequency reaches into the noise band.  Thus a transition IMF is produced, containing signal that has been prematurely removed.  Because this portion of signal no longer remains in the residual, it cannot be accounted for in the next IMF.  Therefore, subsequent IMFs will be damaged as each is derived from the remaining incomplete residual.  This process continues for IMFs 5 and 6, and the portions of the chirp that leak into the empirically defined bands are removed with the noise in a manner similar to IMF 4.  We see the formation of transition IMFs is consistent with the model presented in figure \ref{heuristic}.\\

Spectral leak is similar to the mode mixing observed by \citeauthor{WuHuang} \shortcite{WuHuang}.  To resolve the mode mixing issue, they introduce EEMD to produce IMFs that represent only one scale of oscillation.  EEMD cleverly uses noise perturbations to force the algorithm to explore all frequencies while not adding too much noise so as to push the algorithm into the spectral leak regime.  Noise is added to the original signal and a standard EMD decomposition is performed.  This is repeated with different noise realizations for a fixed number of times.  The resulting IMFs from each run are then averaged, producing an ``ensemble'' result.  \citeauthor{WuHuang} demonstrate that this is an effective way of eliminating mode mixing even in signals that contain a mild amount of noise.  Our analysis continues this line of thought by examining decompositions of signals with noise of higher levels, as is often encountered in real world data.  It is this noise that causes spectral leak between IMFs and presents a different problem than that solved by EEMD.  Adding more noise to the already contaminated signal will not produce cleaner results.  The realization of the original contaminating noise remains the same over all trials and thus cannot be eliminated through averaging.
For these reasons, our analysis is focused on the standard EMD decomposition of noisy signals.\\

\subsubsection{Phase alignment}
\begin{figure}[H]
\centerline{\psfig{file=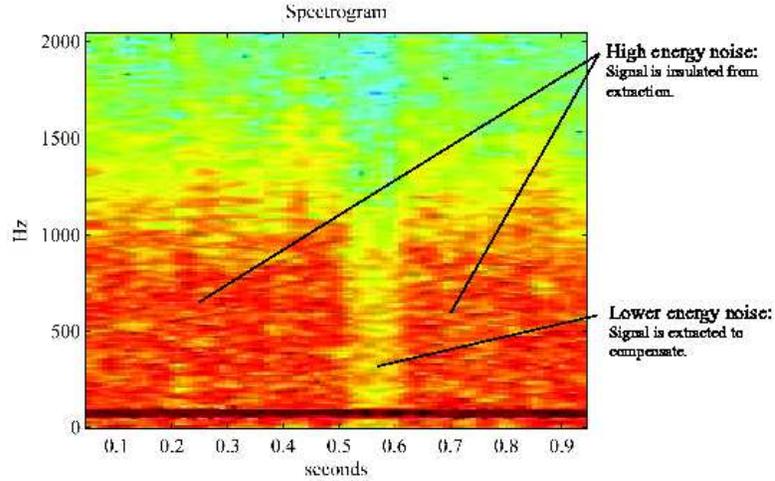,scale=.75}}
\vspace*{8pt}
\caption{A model of a noisy signal in the time-frequency plane.  Signal will be extracted in the region corresponding to 0.5-0.6 seconds.  Here the energy of the noise is too low to insulate the signal from extraction.  Outside of this region, only the energy of the noise will be extracted.}%
\label{phaseHeur}%
\end{figure}

The spectral explanation is not the entire story; the phase of the underlying signal also plays a role in the creation of transition IMFs.  We have seen that the boundaries of the frequency bands of noisy IMFs dip lower in some locations and extend higher in others (figure \ref{noiseIMFsSpec}).  We also have observed that the standard deviation of a band's frequency varies with time (figure \ref{noiseIMFsStat}).  When the energy of the noise is high, the energy of the signal cannot be felt by the algorithm.  In this way, we think of the noise as insulating the signal from extraction.  However, at a given time, if the energy of the noise is small, EMD may include part of the underlying signal in the current IMF as well.  At these time locations, the noise does not insulate the signal from extraction.  Thus signal leaks into an otherwise noisy IMF at the locations where the standard deviation is small.  This process is illustrated by the model seen in figure \ref{phaseHeur}, showing a noisy signal in the time-frequency plane.  From 0.5 to 0.6 seconds there is a clear dip in the energy of the noise.  In this region, the energy of the signal is exposed and will be extracted into the next IMF.  Outside of this region, the energy of the noise is high and insulates the underlying signal.  Here, only the noise will be extracted and the signal will remain untouched.  The locations at which signal is extracted into an otherwise noisy IMF will be shown to be phase dependent.\\

\begin{figure}[h]
\centerline{
\subfigure[$x_1$]{\psfig{file=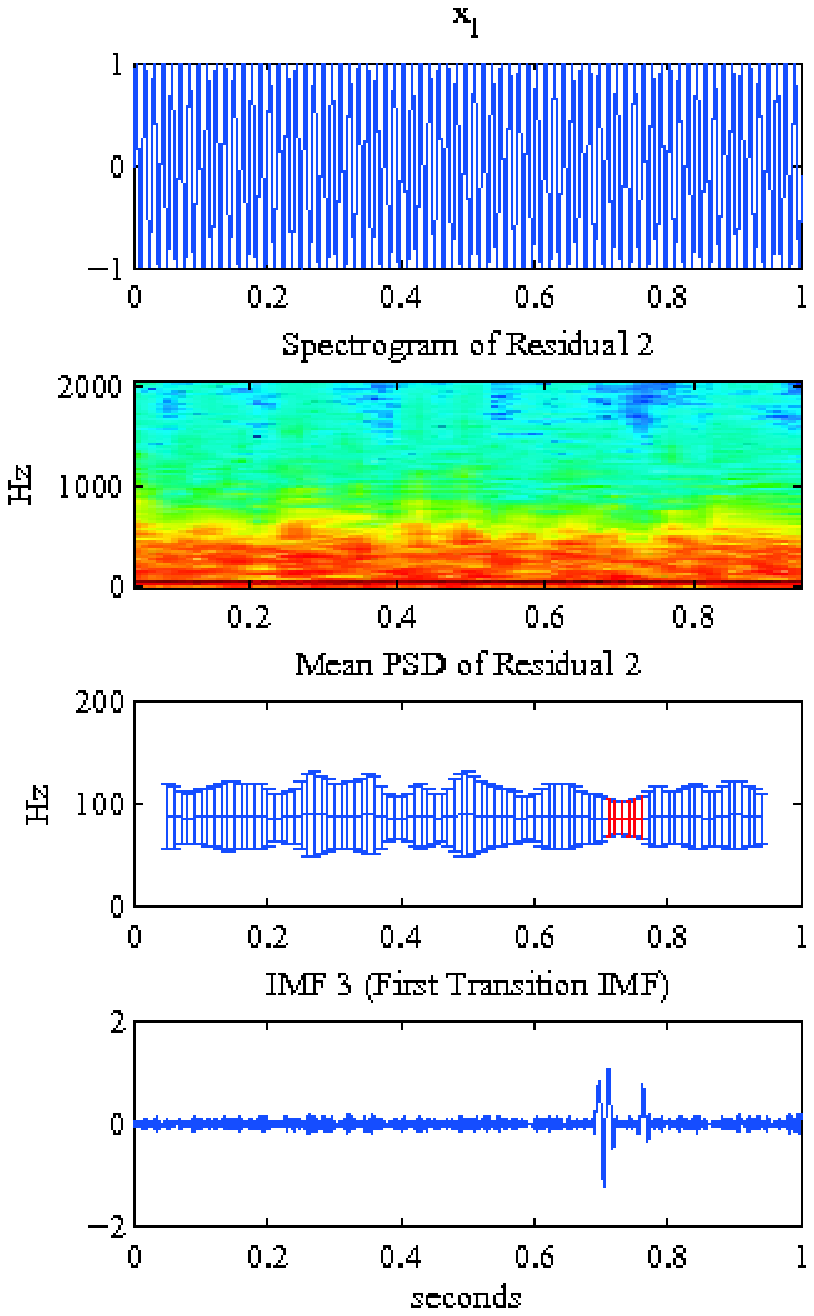,clip,trim=2.5in 2.75in 2.5in 2.75in,scale=.9}}
\subfigure[$x_2$]{\psfig{file=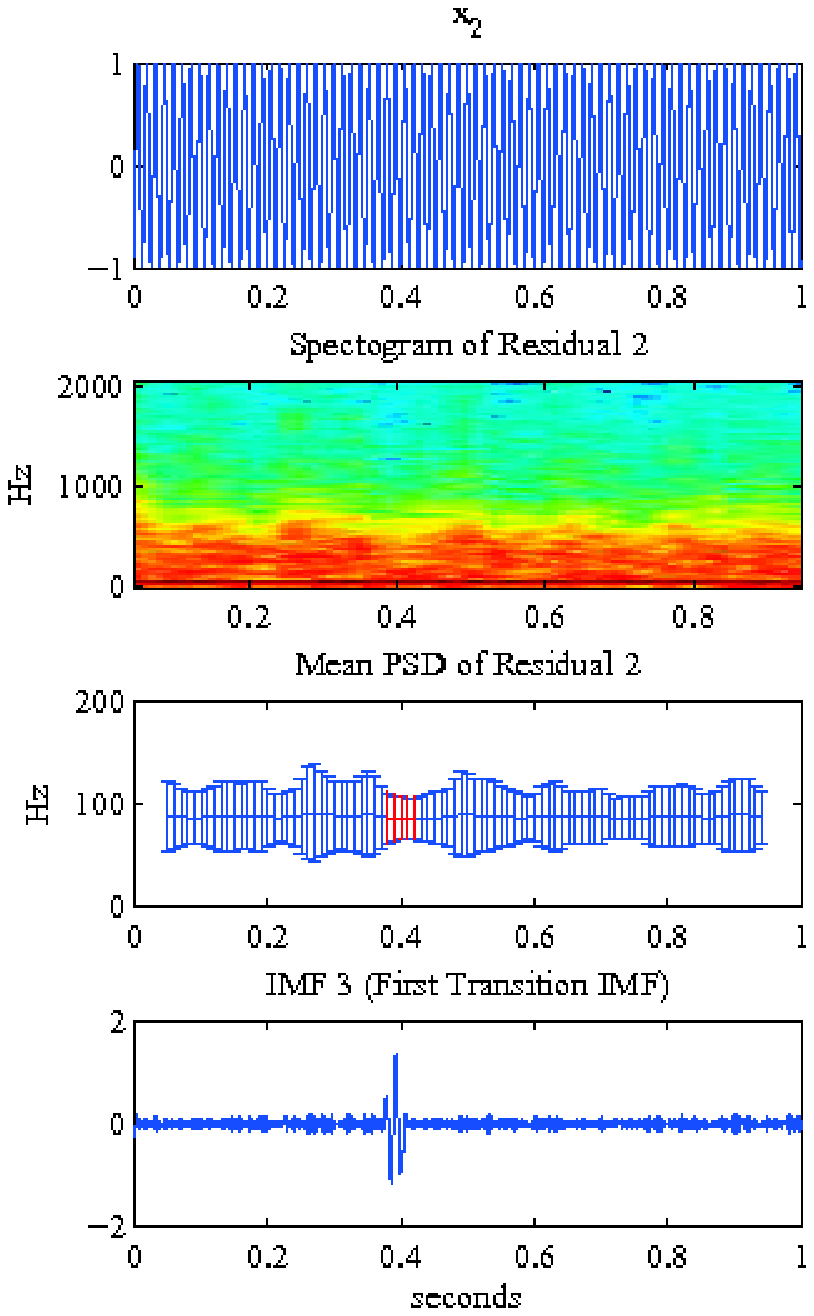,clip,trim=2.5in 2.75in 2.5in 2.75in,scale=.9}}
}
\vspace*{8pt}
\caption{Two stationary signals with identical spectral content differing only by a phase shift.  From top to bottom: the clean signal, spectrograms of the noisy residual from which the first transition IMFs are extracted, mean power spectral density (PSD) of the residual with error bars representing one standard deviation, and the first transition IMFs.  The mean PSD sections highlighted in red (around 0.7s for $x_1$ and 0.4s for $x_2$) correspond to those with the smallest standard deviations and is where signal leaks into the otherwise noisy IMFs.}
\label{phaseIMFs}
\end{figure}

Consider two signals with identical spectral content, differing only by a constant phase factor and contaminated with the same noise realization.  For simplicity, we consider two stationary signals.  Using a stationary example will limit the effect of spectral leak, as unlike the chirp used in the previous nonstationary case, a signal with one frequency should not have energy spread over many IMFs.  Let $f = 75$ Hz and $t \in [0,1]$ seconds.  We examine $x_1 = \sin(2\pi ft)$ and a phase-shifted copy $x_2 = \sin(2\pi ft+.9p)$, where $p = \frac{1}{f}$ is the period of $x_1$.  Because $x_1$ and $x_2$ have the same frequency content, we expect that when contaminated with the same noise realization, EMD should produce very similar results.  Figure \ref{phaseIMFs} shows that the first transition IMFs for each noisy signal contain signal in different locations.  Examining the residual from which each transition IMF was extracted lends an explanation.  The smallest standard deviation in each residual occurs near 0.7 seconds and 0.4 seconds for $x_1$ and $x_2$ respectively and is highlighted in red.  These time locations correspond exactly with the location of signal content in each transition IMF.  At these locations, the level of the noise is too small to insulate the signal from extraction into the current IMF.  This process demonstrates that the extraction of transition IMFs is also phase dependent.\\

In the above example, the first IMF contains pure noise for both signals.  Because the exact same noise realization was used to contaminate both signals, one might expect that the first IMF, and thus the first residual, for each signal would be identical.  However, as noted above and seen in figure \ref{phaseIMFs}, the statistics of the residuals are different, showing dips in the energy of the noise at different locations.  For a more complete understanding of the demonstrated phase dependence, we consider how the phase of a signal interacts with noise.
The interference between the sinusoidal function $x_i(t) = \alpha \cos
(\omega t+ \beta_i)$ ($i=1$ or 2) and a realization $n(t)$ of the
white noise can be described by the following simple model. We
consider $n(t)$ to be a realization of a white noise process sampled
at a finite number of samples $N$.  We can decompose $n(t)$ using
a finite Fourier transform \cite{Brillinger01} and
the Fourier series expansion can be written as follows:
\begin{equation*}
n (t) =  \sum_{k=0}^{N-1}
\rho_k \cos (2\pi k \frac{t}{N}  + \varphi_k)
\end{equation*}
where the $\rho_k \ge 0$ and $\varphi_k$ are defined by
\begin{equation*}
a_k = \rho_k cos \varphi_k, \qquad
b_k = -\rho_k sin \varphi_k, \qquad \text{and} \qquad
a_0 = 2\rho_0 cos \varphi_0,
\end{equation*}
with
\begin{equation*}
a_k = \frac{2}{N} \sum_{t=0}^{N-1} n(t) \cos (2 \pi k\frac{t}{N})  \quad (k =0, \ldots) \qquad
\text{and} \qquad
b_k = \frac{2}{N} \sum_{t=0}^{N-1} n(t) \sin (2 \pi k\frac{t}{N})  \quad (k =1, \ldots).
\end{equation*}
We now contaminate the signal $x_i(t)$ by adding the noise realization
$n(t)$ to $x_i(t)$,
\begin{equation*}
x_i (t) + n (t) =  \alpha \cos
(\omega t+ \beta_i) + \sum_{k=0}^{N-1}
\rho_k \cos (2\pi k \frac{t}{N}  + \varphi_k) \qquad (t=0,1,\ldots, N-1).
\end{equation*}
Because the noise is white, we expect the realization of the noise to
have a uniform distribution of the energy in the Fourier domain. In
other words, we expect that all $\rho_k$ have similar amplitudes.\\

We now examine under what circumstances the noise will interfere with
the signal. First, we assume that the signal amplitude is about the same as the noise level,
($\alpha \approx \rho_{k_0} $). Second, we consider the frequency
index of the noise that matches the frequency of the signal,
$k_0$ such that $\omega \approx 2 \pi k_0$. At this frequency
the noise will interfere with the signal. Formally, we can consider
the interaction of the two cosine function,
\begin{equation*}
\begin{split}
\alpha \cos \left( \omega \frac{t}{N} + \beta_i \right)
& + \rho_{k _0} \cos \left ( 2\pi k_0 \frac{t}{N}  + \varphi_{k_0}
\right) \approx \\
& 2\rho_{k _0}
\cos \left (
\frac{\omega + 2\pi k_0}{2}  \frac{t}{N}  +
\frac{\beta_i+ \varphi_{k_0}}{2}
\right)
\cos \left (
\frac{\omega - 2\pi k_0}{2}  \frac{t}{N}  +
\frac{\beta_i- \varphi_{k_0}}{2}
\right).
\end{split}
\end{equation*}
If $\omega \approx 2 \pi k_0$, then the function
\begin{equation*}
\cos \left (
\frac{\omega - 2\pi k_0}{2}  \frac{t}{N}  +
\frac{\beta_i- \varphi_{k_0}}{2}
\right)
\end{equation*}
slowly modulates the other cosine function,
\begin{equation*}
\rho_{k _0}
\cos \left (
\frac{\omega + 2\pi k_0}{2}  \frac{t}{N}  +
\frac{\beta_i+ \varphi_{k_0}}{2}
\right)
\end{equation*}
which still oscillates at the frequency $\omega$ since $(\omega + 2\pi
k_0)/2 \approx \omega$. The overall amplitude of the slowly varying
envelope $\cos ((\omega - 2\pi k_0)/2 \;t/N + (\beta_i-
\varphi_{k_0})/2)$ clearly depends on the phase difference $(\beta_i-
\varphi_{k_0})/2$, as is shown in figure \ref{phaseIMFs}.\\

We conclude that the exact amount of cancellation created by the
interference between the original signal $x_i(t)$ and the noise
realization $n(t)$ depends on the phase of the signal $x_i(t)$. We
note that this analysis is concerned with one realization of the
noise, and is not in contradiction with the fact that the noise
statistical properties are translation invariant, since the noise is
considered to be stationary.

\section{EMD Decomposition of Synthetic Seismic Data}

Having demonstrated both the effect and mechanism of noise corruption on simple synthetic examples, we turn our attention to a synthetic seismic signal which will serve as a model for real world data.  The signal was constructed using elementary chirplet wave packets.  Such chirplet packets were proposed by \citeauthor{chirplet} \shortcite{chirplet} to decompose seismograms.  Details of the construction are given in Appendix A.  Figure \ref{seismic clean}a shows the clean signal that will be considered along with the estimate of its instantaneous frequency\footnotemark.  In the absence of noise we observe that the decomposition of the signal yields a physically meaningful IF (figure \ref{seismic clean}b).\\
\footnotetext{This synthetic seismic waveform is the result of the superposition of several signals, each with different frequency and amplitude functions.  Therefore, the waveform is a multicomponent signal and its analytic IF is not well defined.  The IF must be computed numerically (as the weighted sum of the IF from each of its IMFs) as shown in figure \ref{seismic clean}b.}
\begin{figure}[h]
\centerline{
\subfigure[Signal]{\psfig{file=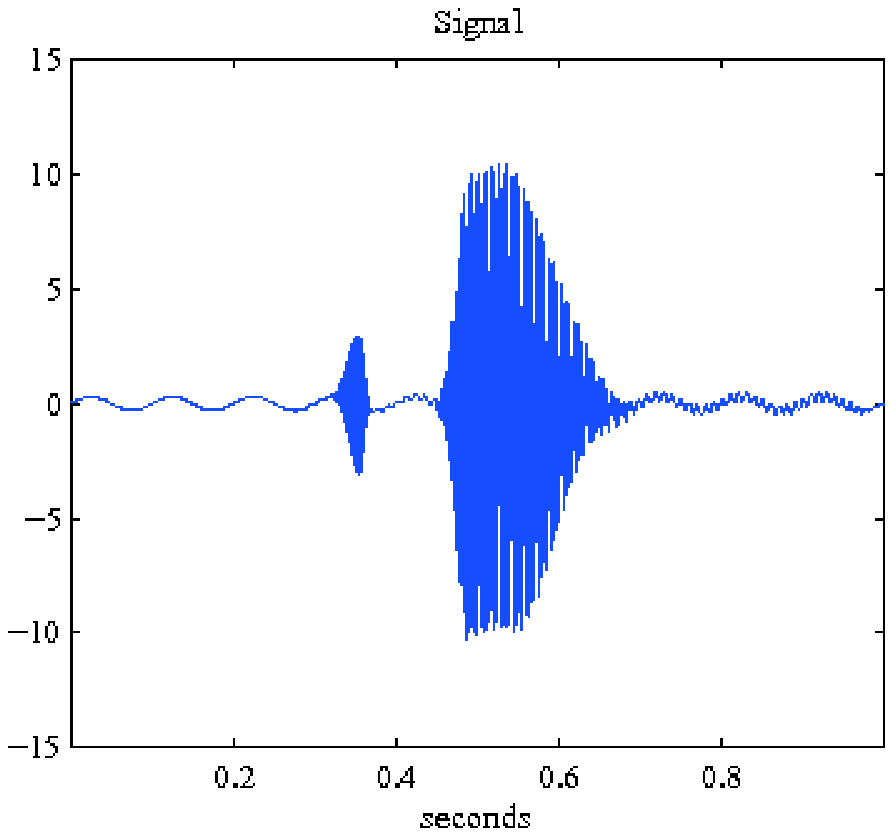,clip,trim= 2.1in 3.5in 2.5in 3.75in,scale=.75}}
\subfigure[Instantaneous Frequency]{\psfig{file=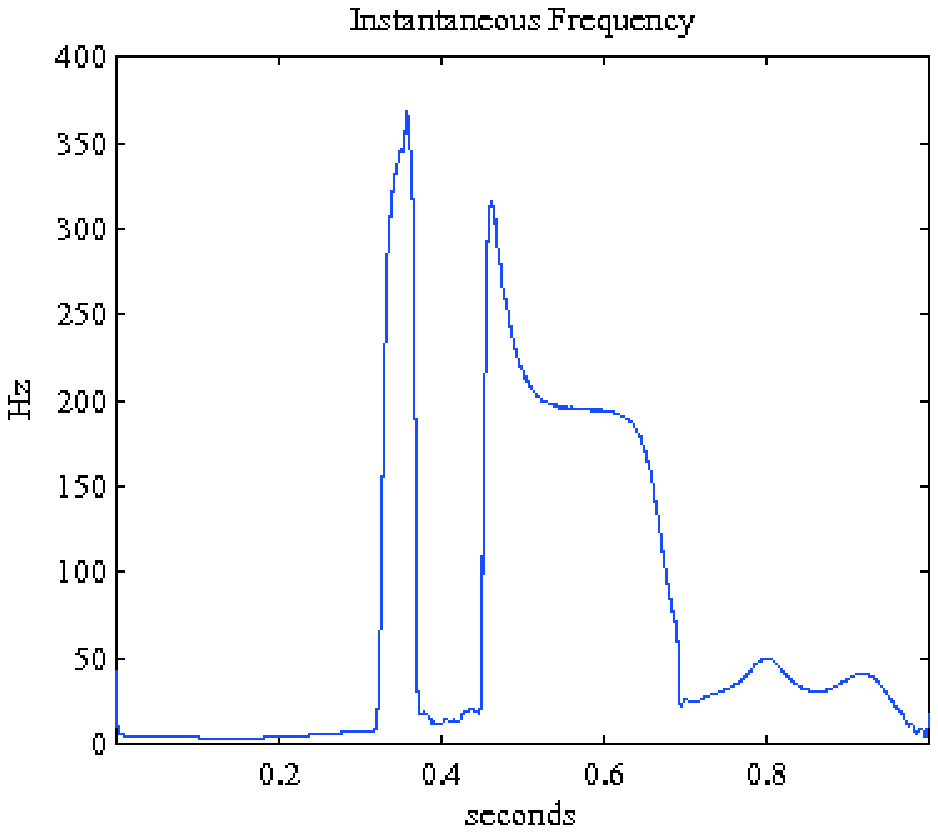,clip,trim= 2.1in 3.5in 2.5in 3.75in,scale=.75}}
}
\vspace*{8pt}
\caption{Clean seismic signal from which a physically meaningful IF is calculated.}
\label{seismic clean}
\end{figure}
\begin{figure}[H]
\centerline{
\subfigure[Signal]{\psfig{file=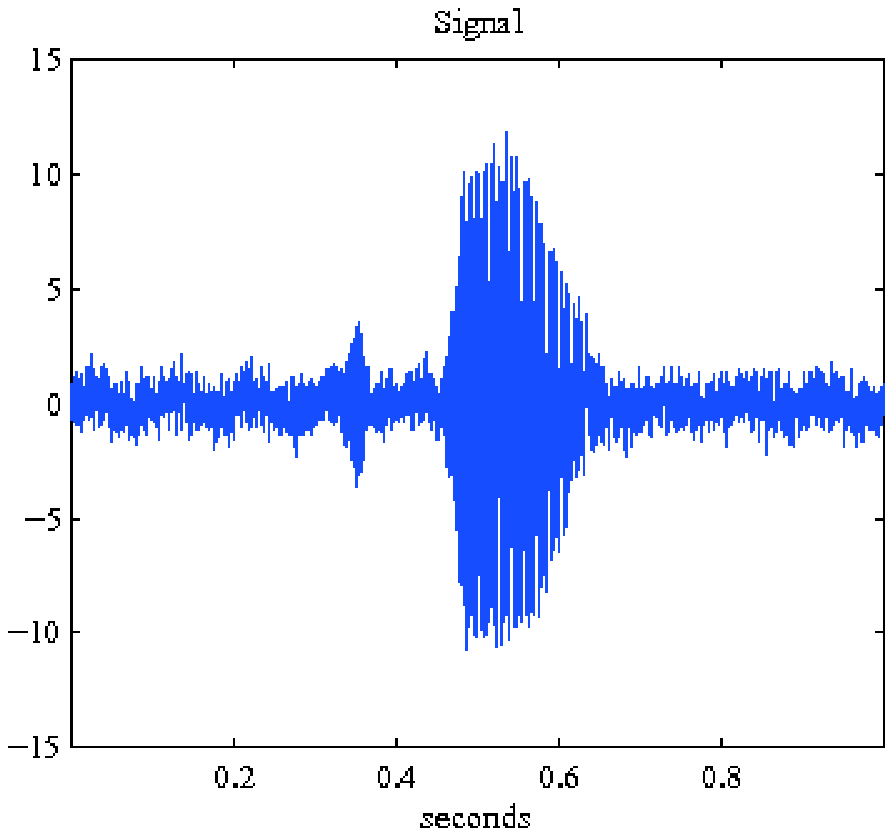,clip,trim= 2.1in 3.5in 2.5in 3.75in,scale=.75}}
\subfigure[Instantaneous Frequency]{\psfig{file=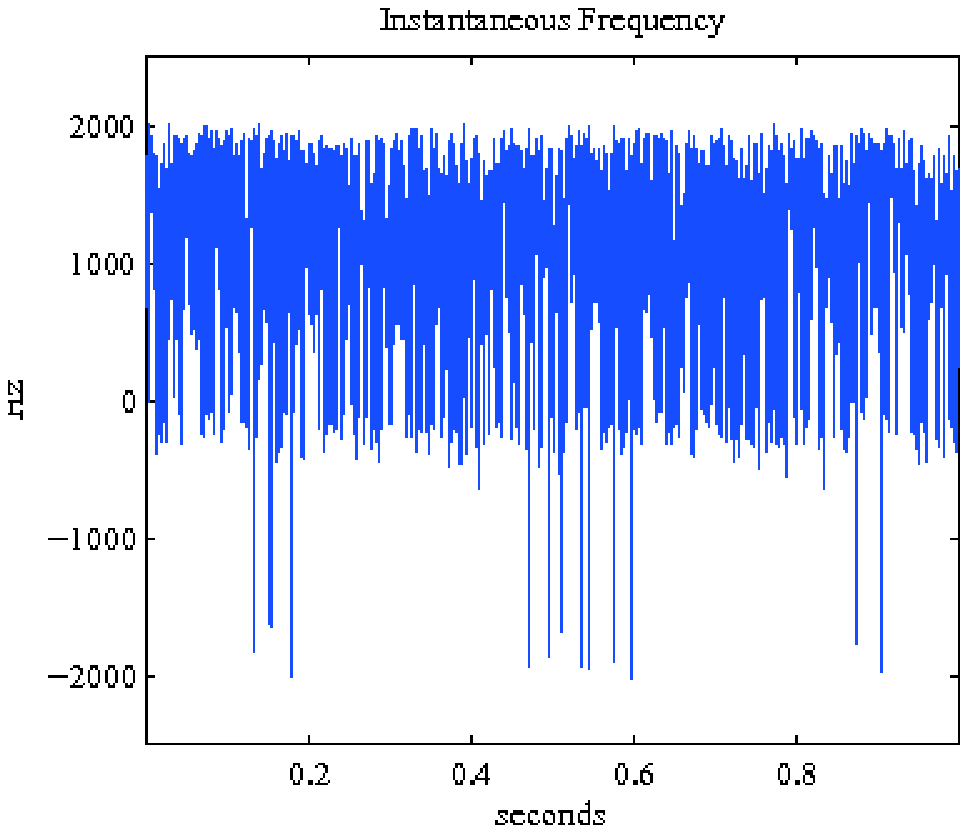,clip,trim= 2.1in 3.5in 2.5in 3.75in,scale=.75}}
}
\vspace*{8pt}
\caption{Noisy seismic signal (SNR = 24dB) from which a physically meaningful IF cannot be calculated.}
\label{seismic noisy}
\end{figure}

\begin{figure}[H]
\centerline{
\subfigure[IMFs]{\psfig{file=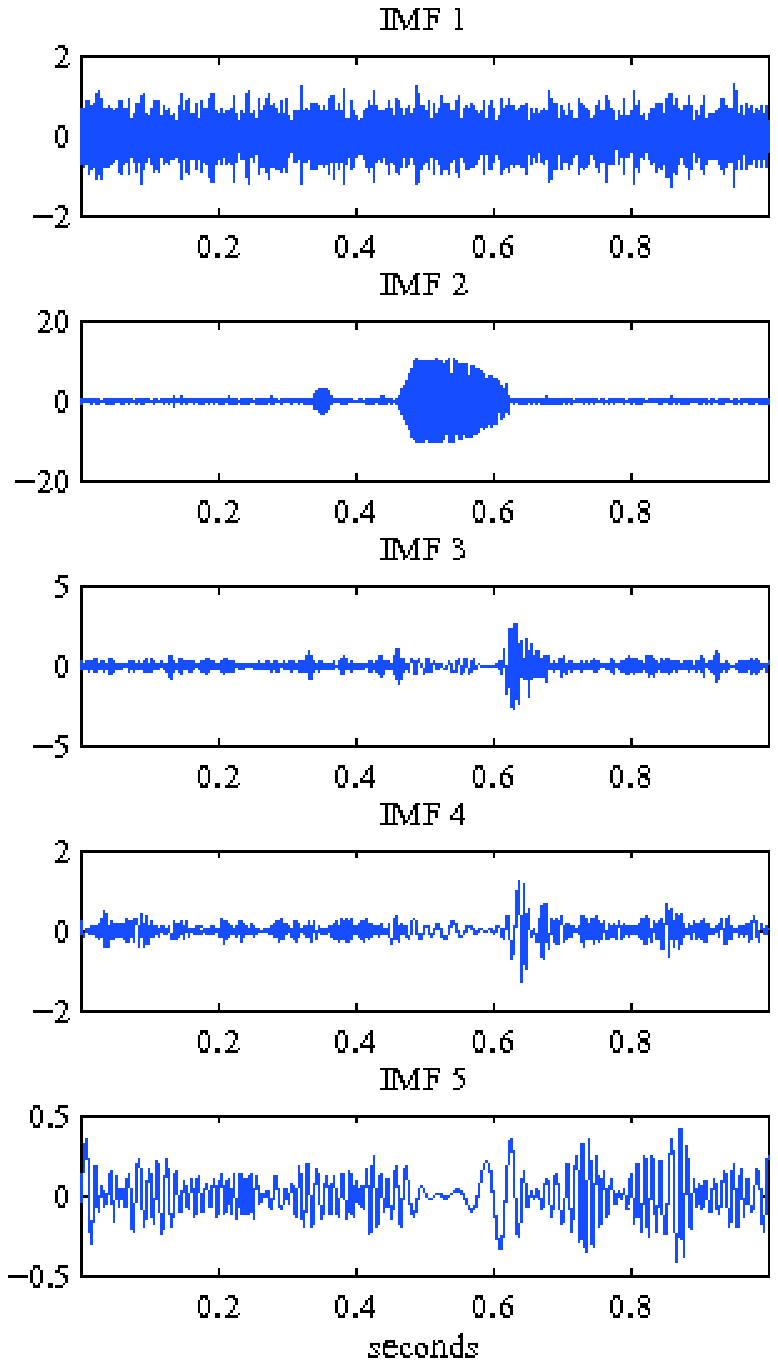,clip,trim=2.5in 2.5in 2.5in 2.75in,scale=.7}}
\subfigure[Spectrograms of IMFs]{\psfig{file=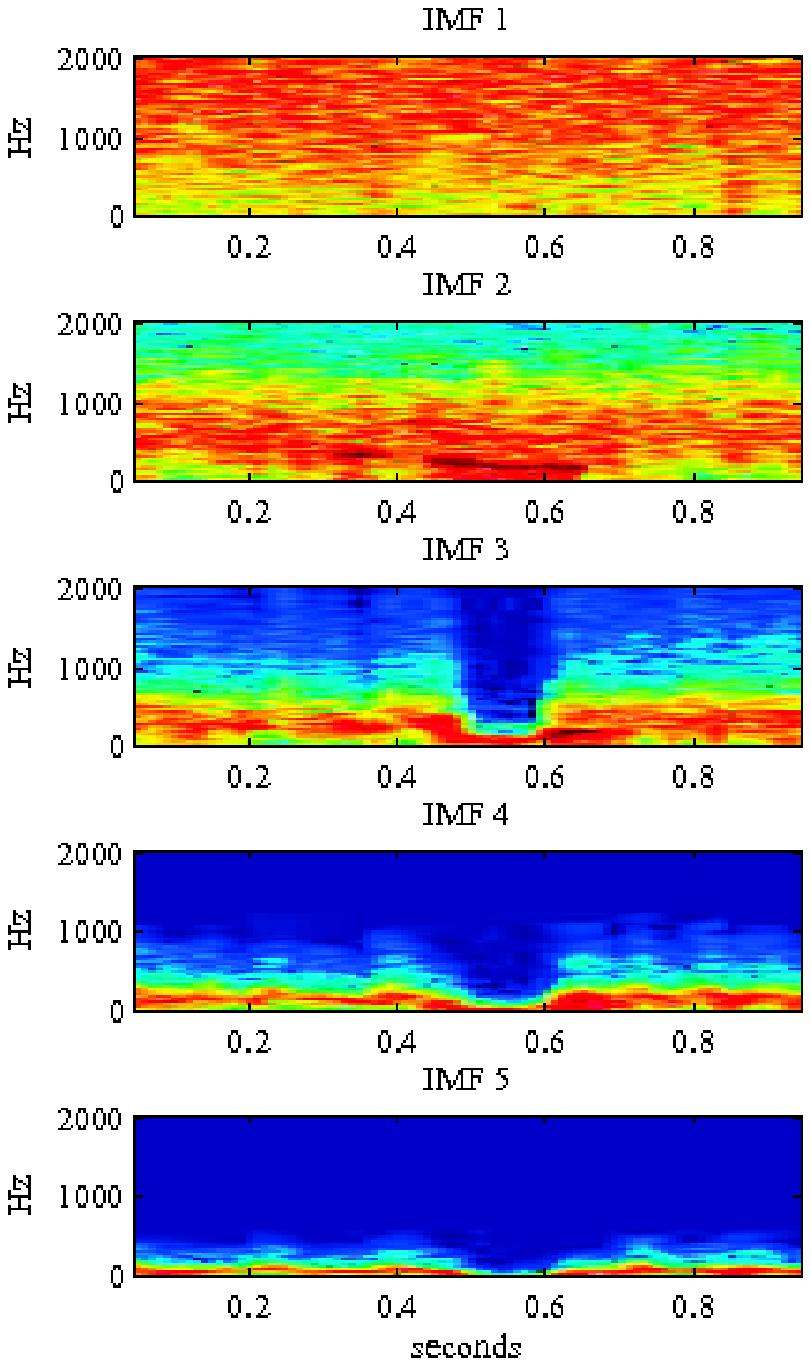,clip,trim=2.5in 2.5in 2.5in 2.75in,scale=.7}}
}
\vspace*{8pt}
\caption{First five IMFs with spectrograms from the decomposition of the noisy seismic signal.  91.8\% of the total energy is captured in transition IMF 2.  IMFs 3-5 are damaged by the extraction of signal into IMF 2.}
\label{seismic noisy IMFs}
\end{figure}

To investigate the effect of noise, the same signal is contaminated with additive white Gaussian noise and we consider an SNR of 24dB.  The noisy signal is shown in figure \ref{seismic noisy}a and it is clear that a meaningful IF was not produced (figure \ref{seismic noisy}b).
Examining the IMFs of the noisy signal shows that IMF 1 contains noise and IMF 2 represents the transition from noise to signal.  It is noted that 91.8\% of the signal's total energy is captured in this transition IMF.  Eleven IMFs were produced and figure \ref{seismic noisy IMFs} shows the first five, capturing 98.6\% of the energy.  It is clear that to produce a meaningful instantaneous frequency, IMF 1 must be discarded.  IMF 2 must be included as it contains almost all of the energy, but will be problematic as it also contains noise.
Recomputing the IF (not shown) using all but the first IMF fails to produce a meaningful IF estimate due to the noise present in IMF 2.\\

The seismic signal is clearly nonstationary.  We therefore expect that the transition IMF was formed due to spectral leak.
The IMFs in figure \ref{seismic noisy IMFs} indicate that the decomposition indeed followed the process presented in the model for spectral leak (figure \ref{heuristic}).  IMF 1 is pure noise, extracted by EMD operating in the filter bank regime.  The spectrogram of IMF 2 shows that EMD continued down the frequency axis in a somewhat dyadic fashion.  In principle, IMF 2 would have contained only pure noise, but the frequency content of the signal leaked into the bottom of this frequency band.  The spectrograms of IMFs 3 - 5 show that the extraction of signal into the transition IMF damaged all subsequent IMFs.\\

\begin{figure}[H]
\centerline{
\subfigure[$x$]{\psfig{file=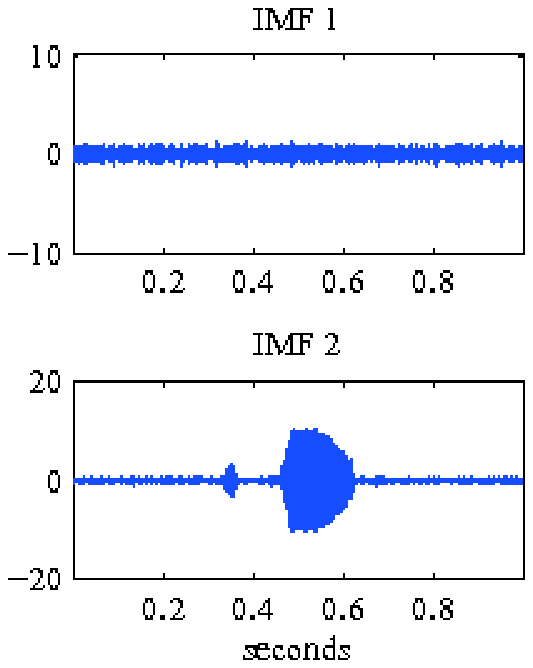,clip,trim=3.1in 4in 3.2in 4.1in,scale=.8}}
\subfigure[$x_1$]{\psfig{file=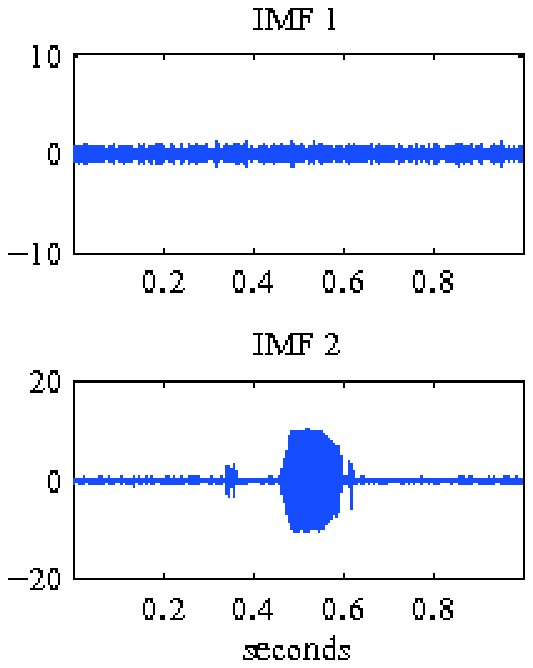,clip,trim=3.1in 4in 3.2in 4.1in,scale=.8}}
\subfigure[$x_2$]{\psfig{file=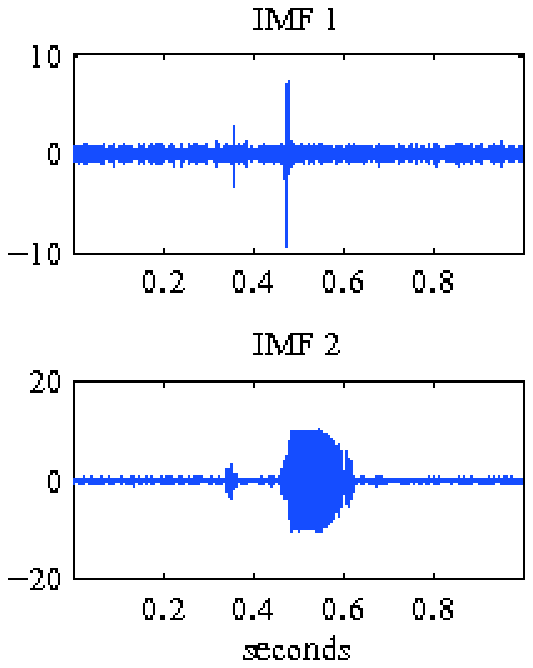,clip,trim=3.1in 4in 3.2in 4.1in,scale=.8}}
}
\vspace*{8pt}
\caption{First two IMFs of noisy seismic signals differing only by a phase factor.  IMF 2 is the transition IMF for $x$ and $x_1$, while the transition begins in IMF 1 for $x_2$.  The transition IMFs for $x$ and $x_1$ contain signal content in slightly different locations, most notable at time $t = 0.6$ seconds.}
\label{seismic phase}
\end{figure}

Finally, there is also evidence of phase dependence.  Let the original signal be denoted by $x$, and consider $x_1$ and $x_2$, two phase-shifted copies of $x$ with identical spectral content.  Phase shift is accomplished by adding a constant $c$ to the argument of the sine in the wave packet $w_k(t)$ (see Appendix A).  The values used for $c$ are $0.9\pi$ and $0.3\pi$ for $x_1$ and $x_2$, respectively.  Figure \ref{seismic phase} shows the transition from noise to signal is captured in IMF 2 for $x$ and $x_1$.  Although subtle, these IMFs contain signal at different locations (most easily seen at 0.6 seconds).  A more obvious effect is seen in the decomposition of $x_2$, where the transition begins in IMF 1 instead of IMF 2.\\

\section{Conclusions}

All data analysis tools are susceptible to noise corruption; EMD is not an exception.  Despite this reality, EMD has emerged as an effective tool for nonstationary data analysis.  Wavelet decompositions, which suffer from similar corruption in the presence of noise, are accompanied by rich theory from which this noise corruption may be studied and understood.  A complete theoretical framework for EMD has yet to emerge.  Therefore, EMD is best understood through experiments to discover and test its limits.  EMD is an effective tool for estimating the IF of a clean signal but provides a poor estimate in the presence of noise.  When decomposing a noisy signal, ``transition'' IMFs are extracted, capturing both noise and signal in the same mode.  Such IMFs are problematic as their noise pollutes the IF calculation yet their signal content cannot be ignored.  We have demonstrated both the existence of and mechanism by which transition IMFs are created.  Specifically, transition IMFs arise from spectral leak between modes and EMD's filter bank behavior in the presence of noise.  In addition, the manner in which signal leaks into an otherwise noisy IMF has been shown to be phase dependent.  Given this understanding, there is an opportunity to more faithfully estimate instantaneous frequency in the presence of noise.  In doing so, care must be taken to treat transition IMFs in a manner that preserves any meaningful physical information, as this is an idea at the core of the development of EMD.\\

\appendix

\section{Seismic Waveform}

The synthetic seismic waveform, $f(t)$, used in section 5 is based on the work of \citeauthor{chirplet} \shortcite{chirplet} and is constructed as follows:\\

\noindent Let $f(t) = \sum_{k=1}^4 a_k w_k\left((t-t_{k})/d_k\right),\quad t\in[0,1]$
\begin{itemize}
	\item Wave packet $w_k(t) = g(t)\,\sin\left[2\,\pi(f_k + p_{k}t^{q_{k}})\,t\right]$
	\item Envelope $g(t)$ = two Gaussians smoothly glued:
	\begin{equation*}
	\begin{cases}
	\exp\left[-\left(\frac{c_k(1-l_k)-t}{\frac{1}{2}c_k(1-l_k)}\right)^2\right]
	& \text{if $0 < t < c_k(1-l_k)$}\\
	1 & \text{if $c_k(1-l_k) < t<c_k+(1-c_k)l_k$}\\
	\exp\left[-\left(\frac{c_k+(1-c_k)l_k-t}{\frac{1}{2}(1-c_k)(1-l_k)}\right)^2\right]
	& \text{if $c_k+(1-c_k)l_k<t<1$}
	\end{cases}
	\end{equation*}
\item $(f_k, p_k, q_k)$ control the frequency of the wave packet
\item $(c_k,l_k)$ control the  boundary between the attack
 and the silencing of the wave packet \newline
\end{itemize}



\noindent The parameter values used are shown in table \ref{tab1} below.\\

\begin{table}[H]
\tbl{Parameters used for construction of seismic waveform.\label{tab1}}
{\begin{tabular}{c c c c c c c c c}
\toprule
$k$ & $t_{k}$ & $d_k$ &$f_k$ & $a_k$& $c_{k}$ & $l_{k}$ & $p_{k}$ & $q_{k}$\\\hline
1 &0 & $1$ & 10 & 0.3 & 0.0 & 1 & 0 &  0 \\
2 & 0.2 & 0.8 & 80 & 0.2 & 0.9 & 0.5 & 10 & 1 \\
3 & 0.32 & 0.05 & 300 & 3 & 0.7 & 0.1 & 2 & -1 \\
4 & 0.45 & 0.24& 195 & 10 & 0.2 & 0.2 & -5& 10 \\
\botrule
\end{tabular}}
\end{table}
\vspace*{12pt}

\section*{Acknowledgments}
This work was supported by the National Science Foundation, project ECS-0501578.  The authors are very grateful to the reviewers for their comments and suggestions.


\begin{thebibliography}

\bibitem[\protect\citeauthoryear{Bardainne {\it et~al}.}{2006}]{chirplet}
Bardainne, T., {\it et~al.} (2006). Characterization of seismic waveforms and classification of seismic events using chirplet atomic decomposition. {E}xample from the {L}acq gas field ({W}estern {P}yrenees, {F}rance). {\it Geophys. J. Int.},
{\bf 166}: 699--718.

\bibitem[\protect\citeauthoryear{Boashash}{1992}]{Boashash1}
Boashash, B. (1992). Estimating and interpreting the instantaneous frequency of a signal {P}art 1: {F}undamentals. {\it P. IEEE.},
{\bf 80}: 520--538.

\bibitem[\protect\citeauthoryear{Brillinger}{2001}]{Brillinger01}
Brillinger, D.R. (2001). {\it Time Series: Data Analysis and Theory}, SIAM.

\bibitem[\protect\citeauthoryear{Flandrin and Goncalves}{2004}]{Flandrin}
Flandrin, P. and Goncalves, P. (2004). The {H}ilbert spectrum via wavelet projections. {\it Int. J. Wavelets Multi.},
{\bf 2}: 477--496.

\bibitem[\protect\citeauthoryear{Huang {\it et~al}.}{1998}]{HuangOrig}
Huang, N.E., {\it et~al.} (1998). The {E}mpirical {M}ode {D}ecomposition and the {H}ilbert spectrum for nonlinear and non-stationary time series analysis. {\it Proc. R. Soc. Lond. A.},
{\bf 454}: 903--995.

\bibitem[\protect\citeauthoryear{Huang {\it et~al}.}{2009}]{onIF}
Huang, N.E., {\it et~al.} (2009). On Instantaneous Frequency. {\it Adv. Adap. Data Anal.}, {\bf 1}: 177--229.

\bibitem[\protect\citeauthoryear{Huang and Shen}{2005}]{HHT}
Huang, N.E. and Shen, S.S.P. (2005). {\it {H}ilbert-{H}uang Transform and Its Applications}, World Scientific.

\bibitem[\protect\citeauthoryear{Kijewski-Correa and Kareem}{2007}]{ND}
Kijewski-Correa, T.L. and Kareem, A. (2007). Performance of wavelet transform and {E}mpirical {M}ode {D}ecomposition in extracting signals embedded in noise. {\it J. Eng. Mech-ASCE.},
{\bf 133}: 849--852.

\bibitem[\protect\citeauthoryear{Loughlin and Tacer}{1997}]{LoughlinTacer}
Loughlin, P.J. and Tacer, B. (1997). Instantaneous frequency and the conditional mean frequency of a signal. {\it Signal Process.},
{\bf 60}: 153--162.

\bibitem[\protect\citeauthoryear{Olhede and Walden}{2004}]{Olhede}
Olhede, S. and Walden, A.T. (2004). {E}mpirical {M}ode {D}ecomposition as data-driven wavelet-like expansions. {\it Proc. R. Soc. Lond. A.},
{\bf 460}: 955--975.

\bibitem[\protect\citeauthoryear{Wu and Huang}{2009}]{WuHuang}
Wu, Z. and Huang, N.E. (2009). {E}nsemble {E}mpirical {M}ode {D}ecomposition: a noise-assisted data analysis method. {\it Adv. Adap. Data Anal.},
{\bf 1}: 1--41.

\end{thebibliography}

\end{document}